\documentclass[usenatbib]{mnras}

\usepackage{amssymb}

\usepackage{newtxtext,newtxmath}

\usepackage[T1]{fontenc}

\DeclareRobustCommand{\VAN}[3]{#2}
\let\VANthebibliography\thebibliography
\def\thebibliography{\DeclareRobustCommand{\VAN}[3]{##3}\VANthebibliography}

\usepackage{graphicx}	
\usepackage{amsmath}	
\usepackage{amssymb, bm}	
\usepackage{color}
\usepackage{multirow}
\usepackage{float}

\usepackage{lineno}

\newcommand{\Rtwohc}{R_{\rm 200c}}
\newcommand{\Mtwohc}{M_{\rm 200c}}

\newcommand{\eg}{{\sl e.g.}, }        
        
\newcommand{\rtwoh}{R_{\rm 200c}}

\newcommand{\Mstarcen}{M_{\star,\, \rm cen}}
\newcommand{\mstarcen}{M_{\star,\, \rm cen}}

\newcommand{\msol}{\ensuremath{\, {\rm M}_\odot}}    
\newcommand{\msun}{\ensuremath{\, {\rm M}_\odot}} 
\newcommand{\kpc}{\ensuremath{\, {\rm kpc}}}

\newcommand{\gyr}{\ensuremath{\, {\rm Gyr}}}

\newcommand{\sigmaDM}{\ensuremath{\sigma_{\rm DM,\,3D}}}
\newcommand{\sigmaDMOneD}{\ensuremath{\sigma_{\rm DM,\,1D}}}
\newcommand{\sDM}{\ensuremath{s_{\rm DM}}}
\newcommand{\sDMvel}{\ensuremath{s_{\rm DM,\,vel}}}
\newcommand{\Gammadyn}{\ensuremath{\Gamma_{\rm dyn}}}
\newcommand{\ctwohc}{\ensuremath{c_{\rm 200c}}}
\newcommand{\aform}{\ensuremath{a_{\rm form}}}
\newcommand{\EsDM}{\ensuremath{E_{\rm s,\,DM}}}

\newcommand{\tdyn}{\ensuremath{t_{\rm dyn}}}
\newcommand{\pSHAP}{\ensuremath{\phi_{\rm SHAP}}}
\newcommand{\MBHpiv}{\ensuremath{M_{\rm BH,\,piv}}}

\definecolor{bleudefrance}{rgb}{0.19, 0.55, 0.91}
\definecolor{purple}{RGB}{128, 0, 128}

\newcommand*{\vcenteredhbox}[1]{\begingroup
\setbox0=\hbox{#1}\parbox{\wd0}{\box0}\endgroup}

\newcommand{\OrcidIDName}[2]{\href{https://orcid.org/#1}{#2}}

\title[Baryonic Imprints on DM Halo Population]{Baryonic Imprints on DM Haloes: Population Statistics from Dwarf Galaxies to Galaxy Clusters}

\author[Dhayaa Anbajagane]{\OrcidIDName{0000-0003-3312-909X}{Dhayaa Anbajagane}\thanks{Corresponding author email: dhayaa@uchicago.edu}(\vcenteredhbox{\includegraphics[height=1.2\fontcharht\font`\B]{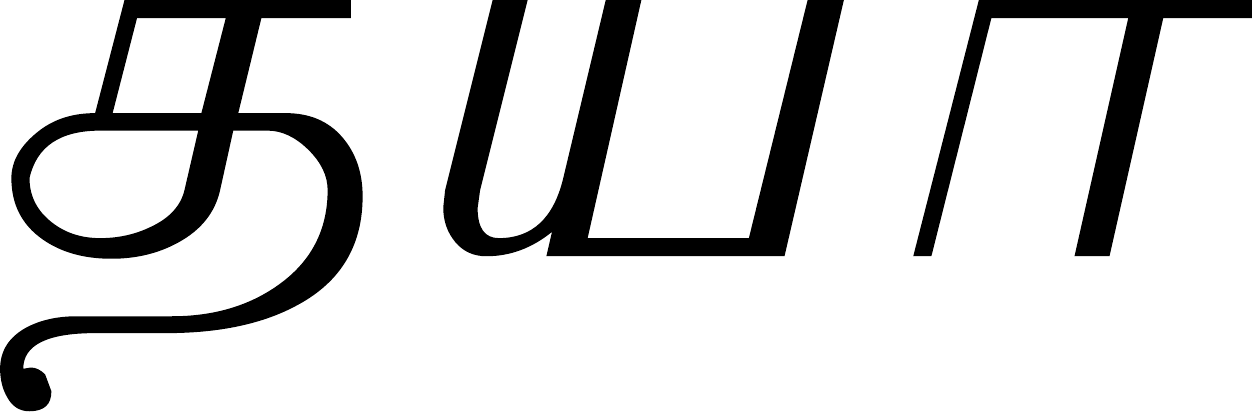}}),$^{1,\, 2}$
\OrcidIDName{0000-0002-4876-956X}{August E. Evrard},$^{2,\, 3}$
\OrcidIDName{0000-0003-0777-4618}{Arya Farahi},$^{4,\, 5}$
\\
\\
$^1$Department of Astronomy and Astrophysics, University of Chicago, 5640 S. Ellis Ave, Chicago, IL 60637, USA\\
$^2$Department of Physics and Leinweber Center for Theoretical Physics, University of Michigan, Ann Arbor, MI 48109, USA\\
$^3$Department of Astronomy, University of Michigan, Ann Arbor, MI 48109, USA\\
$^4$ Michigan Institute for Data Science, University of Michigan, Ann Arbor, MI 48109, USA \\ $^5$ Department of Statistics and
Data Science, The University of Texas at Austin, TX 78712, USA} 
\date{Accepted XXX. Received YYY; in original form ZZZ}

\pubyear{2021}

\begin{document}

\label{firstpage}
\pagerange{\pageref{firstpage}--\pageref{lastpage}}
\maketitle

\begin{abstract}

In a purely cold dark matter universe, the initial matter power spectrum and its subsequent gravitational growth contain no special mass- or time-scales, and so neither do the emergent population statistics of internal dark matter (DM) halo properties. Using $1.5$ million haloes from three \textsc{IllustrisTNG} realizations of a $\Lambda$CDM universe, we show that galaxy formation physics drives non-monotonic features (``wiggles'') into DM property statistics across six decades in halo mass, from dwarf galaxies to galaxy clusters.
We characterize these features by extracting the halo mass-dependent statistics of five DM halo properties --- velocity dispersion, NFW concentration, density- and velocity-space shapes, and formation time --- using kernel-localized linear regression (\textsc{Kllr}). Comparing precise estimates of normalizations, slopes, and covariances between realizations with and without galaxy formation, we find systematic deviations across all mass-scales, with maximum deviations of 25\% at the Milky-Way mass of $10^{12} \msol$. The mass-dependence of the wiggles is set by the interplay between different cooling and feedback mechanisms, and we discuss its observational implications. The property covariances depend strongly on halo mass and physics treatment, but the correlations are mostly robust. 
Using multivariate \textsc{Kllr} and interpretable machine learning, 
we show the halo concentration and velocity-space shape are principal contributors, at different mass, to the velocity dispersion variance.  
Statistics of mass accretion rate and DM surface pressure energy are provided in an appendix.  We publicly release halo property catalogs and \textsc{Kllr} parameters for the TNG runs at twenty epochs up to $z = 12$.
\end{abstract}



\section{Introduction}

Our Universe is populated with galaxies hosted by dark matter (DM) haloes of a wide dynamic mass range. The entirety of this mass range is abundant with information; from the dwarf galaxies whose inner profiles constrain the particle nature of dark matter \citep[see][for reviews]{Bernal2017FIMPReview, Knapen2017LightDMReview, Drlica-Wagner2019WhitePaperLSSTDM}, to regular-type galaxies whose clustering properties allow for leading cosmological analyses \citep[see][for reviews]{Weinberg:2013review, Mandelbaum2018WeakLensingReview}, to galaxy clusters whose number counts strongly trace both the expansion and growth history of the Universe \citep[see][for reviews]{Allen2011CosmoClusterReview,Weinberg:2013review,Huterer:2018review}. While the observational component of cosmological studies is focused on stellar/gas properties, the underlying theory that connects galaxies to cosmology is more closely related to their host \textit{DM haloes}, which are extended distributions of DM within which galaxies form and evolve.

The character of the halo population is thus central to cosmology. There is a rich history of literature on the formation of haloes using cosmological simulations \citep[\eg][]{Bertschinger1999SimReview} as well as on their emergent configurations in a multi-dimensional space of essential properties such as shape (in density- or velocity-space), gravitational potential, accretion rate and merger history, the local large-scale environment, subhalo and galactic content, and more  \citep[see][for reviews]{Cooray2002HaloModel, Kravtsov2012ClusterFormation, Wechsler2018GalaxyHaloConnection}. Some of these properties were first studied by the community more than half a century ago. The pioneering N-body simulation of \citet{Peebles1970Virial}, which followed 300 particles undergoing purely gravitational interactions, showed that the collective relaxation of a bound system happens rapidly; from a cold start, the system achieved virial equilibrium after essentially a single crossing time. As a result, the first scaling relation of particle velocity dispersion with halo mass was analyzed for a cold dark matter (CDM) cosmology \citep{Evrard1989BiasedCDM}, and a larger, meta-analysis of cluster-scale haloes showed this was a particularly tight relation with only $\approx 4\%$ fractional scatter \citep{Evrard2008VirialScaling}. The key discovery by \citet{Navarro1997NFWProfile} that haloes relax toward an internal matter density profile described by a single-parameter --- a concentration value --- was followed by linking a halo's concentration to its assembly history \citep{Wechsler2002Concentrations}. These original halo properties are among the key properties we investigate in this paper.

There has thus been extensive work on the origin of relationships between such DM halo properties (commonly denoted as ``secondary'' properties) and halo mass, both from a theoretical, analytic perspective \citep[\eg][]{Dalal2010OriginsDmHaloProfiles, Okoli2016Concentration}, as well more simulation-based, numerical ones \citep[\eg][]{Allgood2006Shapes, Evrard2008VirialScaling, Ludlow2014MilleniumConcentration, Bonamigo2015MilleniumHaloShapes, Diemer2015Concentration, Vega-Ferrero2017MdplDmoHaloShapes,  Diemer2019concentrations, Ishiyama2020UchuuConcentration, Hearin2021DiffMAH}.
Models for astrophysical and cosmological observables have also frequently utilized the scaling relations of such DM halo properties, and these relations were in turn predominantly calibrated using N-body simulations of a single, collisionless fluid (referred to as dark matter only, or DMO simulations). However, such scaling relations are not the ``true'' result as baryonic processes, which are necessary for modelling a realistic universe, alter the DMO behavior. Thus, it is important to quantify the baryonic imprints on these relations as they will inform how appropriate it is to continue using a DMO-derived scaling relation in a model given the accuracy requirements of the analysis.

In the past decade, increases in computational capacity and capability have enabled the realization of families of large-scale, cosmological hydrodynamics simulations that evolve multiple, interacting fluids representing the DM and baryonic matter components.  Each family aims to capture a plethora of astrophysical processes, such as magneto-hydrodynamics, radiative cooling, stellar and supermassive black hole (SMBH) formation, kinetic and thermal feedback mechanisms via active galactic nuclei and supernovae, chemical enrichment and turbulence. We refer to this class of  cosmological simulations as ``Full Physics'' (FP) treatments.  In this paper we use the \textsc{IllustrisTNG} family of simulations to examine the mass-dependent impact of FP treatments on the DM property scaling relations.

Previous works have probed such differences  \citep[\eg][]{Kazantzidis2004HaloShapes,  Duffy2008Concentration, Stanek2009GasPhysicsMF, Tissera2010BaryonImprints, Pedrosa2010BaryonImprints, Knebe2010Shapes, Bryan2013BaryonImpactOnShapes, Velliscig2015EagleOwlsHaloShapes, Cui2016NiftyBaryonsHaloProperties, Henson2017BaryonEffectsBM, Armitage2018C-EagleVelDisp, Artale2019BaryonImprints, Ragagnin2019HaloConcentration}, but due to the natural compromise between numerical resolution and simulation volume, most were limited to a relatively narrow range in halo mass and focused on a particular astronomical population, such as dwarf galaxies, Milky-Way (MW) type galaxies, or galaxy clusters. Several studies have also gotten around this issue by using multiple simulation realizations, each with the same underlying astrophysical modeling but different resolution and comoving box sizes \citep[\eg][]{Bryan2013BaryonImpactOnShapes, Velliscig2015EagleOwlsHaloShapes, Ragagnin2019HaloConcentration}. This strategy, which widens the available range in halo mass by combining different realizations, is the approach we employ in this work.

Since astrophysical processes --- unlike gravitational ones --- are scale-dependent, we can expect the dark matter halo structure of FP runs to have nonlinear features in its property scaling relations.
We present here measurements of the \textit{halo mass-dependent} impact of baryonic physics on scaling relation parameters (normalizations, slopes, and covariance) for DM halo properties.  Our study spans a wide halo mass range, from dwarf galaxy-scales ($\Mtwohc \sim 10^{9} \msol$) to galaxy cluster-scales ($\Mtwohc \sim 10^{14.5} \msol$), and our key goals are to: (i) measure precise, halo mass-dependent scaling relations for both DMO and FP runs over a wide dynamic range of halo mass, (ii) determine baryon back-reaction\footnote{We use ``back-reaction'' here to refer to the non-linear gravitational coupling between baryons and DM. Other meanings of the term exist in other contexts.} effects, and the impacts of FP treatments on DM halo properties via comparisons with DMO counterparts, and; (iii) characterize the intrinsic property (co)variances, and the ability of other halo properties to reduce this variance, with the latter goal explored using both local linear regression and an interpretable machine learning method. A corollary aim of our analysis is the resolution dependence of the scaling relations in both FP and DMO runs.

This paper is organized as follows: in  \S\ref{sec:Data} we describe the simulations we employ, and in \S\ref{sec:Methods}, we define the halo properties and the methods used in the analysis. We then present our results for the mass-dependent scaling relation parameters in \S\ref{sec:Fit_Params}, including observational prospects for the extracted features, and in \S\ref{sec:explain_scatter}, present the correlation matrix alongside the role of the secondary halo properties in reducing the scatter of the $\sigmaDM - \Mtwohc$ scaling relation. In \S\ref{sec:Discussion}, we summarize some of our main findings alongside discussions of our results, before concluding in \S\ref{sec:conclusions}. Results for additional halo properties of interest are also shown in Appendix \ref{appx:Additional_Properties}, while the redshift dependence is extracted from some properties in Appendix \ref{appx:Redshift_evol}. Our halo mass convention is $\Mtwohc$, defined in \S\ref{sec:mass_def}, and expressed in units of solar masses, $\msol$.

\section{Data} \label{sec:Data}

This work is based on the \textsc{IllustrisTNG} simulations \citep{Springel2018FirstClustering, Pillepich2018FirstGalaxies, Nelson2018FirstBimodality, Marinacci2018FirstFields, Naiman2018FirstEuropium, Nelson2019TNG50, Pillepich2019TNG50}, which are the follow-up to the \textsc{Illustris} project \citep{Vogelsberger2014Illustris}, and are generated using the \texttt{AREPO} moving-mesh code \citep{Springel2010EMesh}. They include an extensive prescription of baryonic physics, such as metal cooling, stellar-formation, and active galactic nuclei (AGN) feedback sourced by supermassive black holes, to name a few. The specific modeling and parameter choices are detailed in \citet{Weinberger2017Methods, Pillepich2018Methods}.

We utilize the DMO and FP variants of the highest resolution TNG50, TNG100 and TNG300 simulations, and Table~\ref{tab:Sim_data} offers information for these runs such as resolution and box volume. Note that the FP and DMO realizations of each run start from the exact same initial conditions, drawn from a Planck 2015 cosmology \citep{Planck2015CosmoParams}. The three FP runs all share the same galaxy formation model configurations as well.

The three volumes have different spatial resolution but their overlap in halo mass scale enables studies of numerical resolution on property scalings. We quote results for each volume within mass ranges given by the right column of Table~\ref{tab:Sim_data}.  
The maximum halo mass is set by the effective halo counts reaching $N_{\rm eff} = 200$ --- see equation \eqref{eqn:KLLR_weights} and Figure \ref{fig:Halo_Counts_effective} below --- while the minimum mass-scale is the mass of a halo that has $\approx\!1500$ particles within $r < \Rtwohc$. We make accommodations in both limits to set the numbers to be multiples of $0.1\,\,\rm dex$. Note that a hard particle-count threshold cannot be jointly defined for both DMO and FP runs in an umambiguous way, as the baryonic elements can have evolving masses unlike the DM particles which have fixed mass. Finally, to prevent edge effects in our local regression method, which is described in \S\ref{sec:KLLR}, the minimum mass of our sample is $0.2\,\, \rm dex$ below that given in Table~\ref{tab:Sim_data}, but we only report results for the mass ranges denoted in the table. For the DMO runs, the total number of haloes above the mass thresholds are as follows: 105500, 125719, and 324884 for TNG 50/100/300 DMO, respectively. The total number of haloes above the \textit{minimum mass} from all six runs --- three each for DMO and FP treatments --- is over 1.5 million. 

\begin{table}
    \centering
    \begin{tabular}{l|c|c|c|c|c}
         & $L_{\rm box} $ & $m_{\rm \,DM}$ || $m_{\rm\, b}$ & $\epsilon_{\,\rm DM}^{z = 0}$ & $\log_{10}M_{\rm KLLR}$ \\[0.5em]
        Run & [Mpc] & $[10^6 \, \msol]$  & [kpc] & $[\msol]$\\
        \hline
    TNG50 & $52$ & $0.45$  || $0.08$ & $0.29$ & $9.0 - 12.5$\\
    TNG100 & $111$ & $7.5$  || $1.4$ & $0.74$ & $10.0 - 13.5$\\
    TNG300 & $303$ & $59$  || $11$ & $1.48$ & $11.0 - 14.5$ \\[0.05cm]
    TNG-MV & $37$ & $12.4$  || $2.4$ &  $0.74$ & $10.5 - 12.8$ \\
        \hline
    \end{tabular}
    \caption{Properties of the FP simulations (left to right): comoving length of simulation box, DM particle mass, target baryon mass, Plummer equivalent gravitational softening, and halo mass range of \textsc{Kllr} statistics. The \textsc{Kllr} statistics are measured using steps of 0.1 dex within these mass ranges. The DMO runs share the same parameters except DM particle mass: $m_{\rm DM}^{\rm DMO} = m_{\rm DM}^{\rm FP} + m_{\rm b}^{\rm FP}$. TNG-MV parameters are common to all the ``model variant'' runs described in section \ref{sec:Model_variants}.  
    }
    \label{tab:Sim_data}
\end{table}

Our specific definitions of halo properties are detailed  in \S\ref{sec:DMO_property_def}.  For reasons discussed below, our measures focus on material associated with the central, or primary, subhalo.

\subsection{TNG Model Variant Runs and SMBH Feedback} \label{sec:Model_variants}

We utilize an extensive library of TNG model variant runs \citep{Pillepich2018Methods} to test the sensitivity of our results to parameters in the TNG galaxy formation model. These runs, with $L_{\rm box} = 37\, \rm Mpc$, start from the same initial conditions, and are run with specific galaxy formation modules turned on/off, or with their tuning parameters varied, in comparison to the fiducial TNG model. In this work, we focus on models that vary parameters associated with SMBHs.

SMBHs in TNG have both a radio mode, with kinetic energy injected into the halo, and a quasar mode, with thermal feedback injected into nearby gas. The transition between the two modes occurs when the ratio of the Bondi-Hoyle-Lyttleton accretion rate \citep{Hoyle1939ISM, Bondi1944StellarAccretion, Bondi1952Accretion} to the Eddington accretion rate crosses a threshold given by,
\begin{equation} \label{eqn:BH_Accretion_Mode}
    \chi = {\rm min}\bigg[0.002\bigg(\frac{M_{\rm BH}}{M_{\rm BH,\,piv}}\bigg)^2, 0.1\bigg].
\end{equation}
SMBH accretion rates above and below $\chi$ produce quasar and radio mode response, respectively.

The energy injection into the gas surrounding the SMBH is a discretized process, happening only after the SMBH accumulates enough energy via accretion. In the radio mode, the gas near the SMBH is injected with kinetic energy in the form of a momentum kick, but the \textit{direction} of the kick is determined randomly. Averaged over time, the distribution of kinetic energy injected by a given SMBH into its host halo is \textit{isotropic}, and a potential factor relevant to our discussion of halo shapes below.

In this work, we utilize TNG variant models \texttt{3403} and \texttt{3404}, where the pivot mass --- whose fiducial value is $M_{\rm BH,\,piv} = 10^8 \msol$ --- is increased and decreased by a factor of $4$ respectively. We show below how nonlinear features in scaling relations respond to the choice of pivot mass scale.

\section{Methods} \label{sec:Methods}

We begin by describing our estimator for the scaling relation parameters. Our local linear regression method and the effective sample size as a function of halo mass are described in \S\ref{sec:KLLR}.  We then define the set of DM properties measured for each halo in \S\ref{sec:DMO_property_def}. The reader eager to reach the results section is encouraged to scan Table \ref{tab:Halo_properties} and Figure~\ref{fig:Halo_Counts_effective}, and refer back here for details as needed.

\subsection{Kernel Local Linear Regression (\textsc{Kllr})} \label{sec:KLLR}

To extract the mass-dependent scaling relations, we utilize the Kernel-Localized Linear Regression (\textsc{Kllr}) method\footnote{\url{https://github.com/afarahi/kllr}}, first described in \citet{Farahi2018BAHAMAS}, but with one key modification detailed further below.

The fiducial method extracts the scaling relation parameters centered at a particular halo mass-scale, $M_c$, by weighting haloes with a Gaussian in $\ln(M/M_c)$ and then performing ordinary, least-squares linear regression on the weighted halo properties to extract normalizations, slopes, and covariance at mass-scale, $M_c$. Throughout this work we use a Gaussian kernel for the weights with $\sigma_{\rm KLLR} = 0.2\ {\rm dex}$ (which is $0.46$ in natural log terms). Systematically incrementing $\ln M_c$ produces effectively-continuous estimates of the scaling relation over a range of mass, with exactly-continuous results achieved in the limit $N_{\rm halo} \rightarrow \infty$ and $\sigma_{\rm KLLR} \rightarrow 0$. 

We note here that the 1500-particle minimum limit means that measurement uncertainties in the global halo properties we study are very small compared to the intrinsic population scatter.  With one exception, we therefore ignore measurement uncertainties during the regression, and interpret the \textsc{Kllr} mass-dependent covariance elements as estimating the \textit{intrinsic population covariance} of these properties. The exception is the halo NFW concentration, $\ctwohc$, and this is discussed further in section \ref{sec:c200c_def} and \ref{sec:c200c_Fit}.

\begin{figure}
    \centering
    \includegraphics[width = \columnwidth]{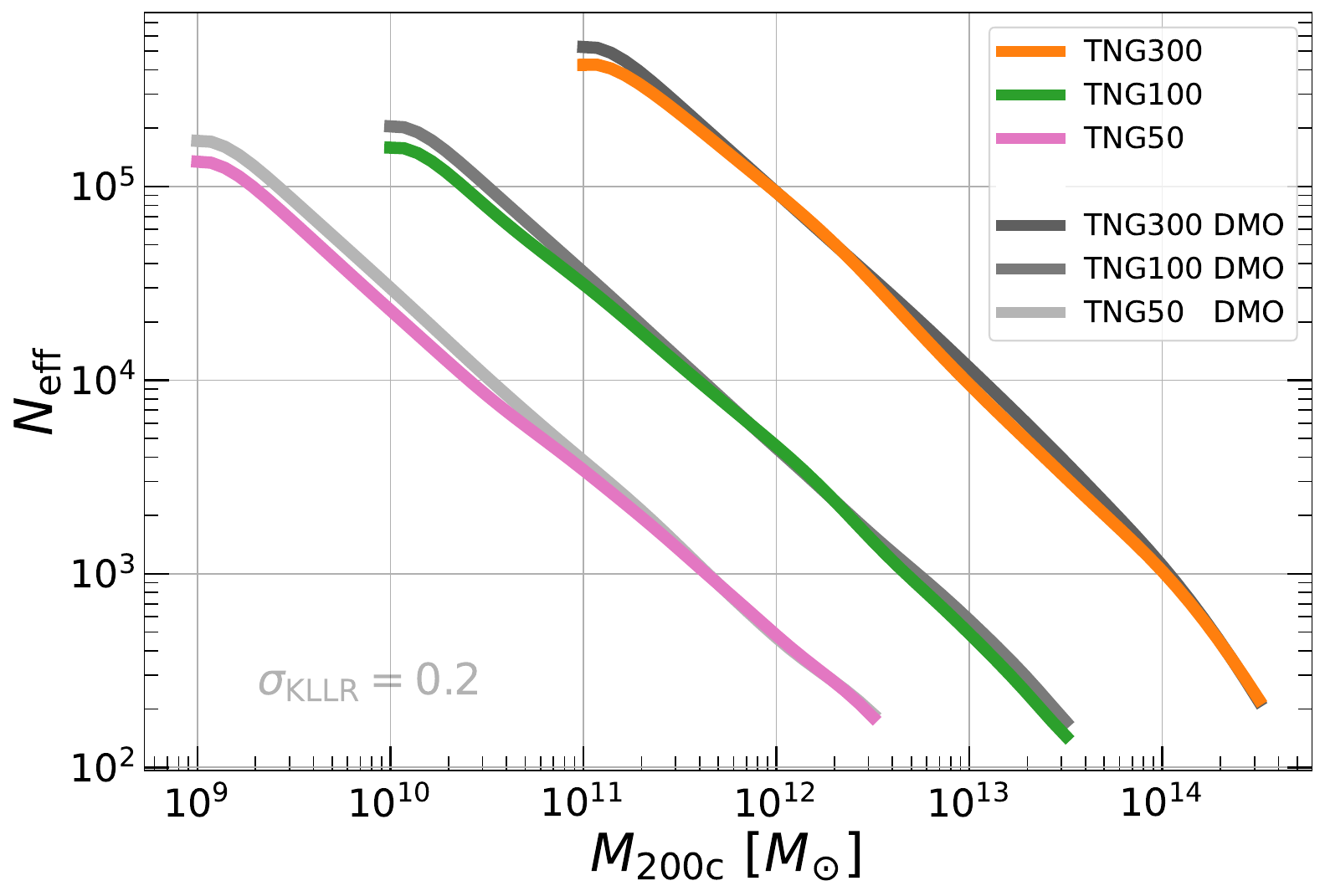}
    \caption{The \textit{effective} number of haloes,  $N_{\rm eff}(M)$, defined by the sum of Gaussian weights, equation~\eqref{eqn:N_halos_effective}, using the kernel width $\sigma_{\rm KLLR} = 0.2\,\, \rm dex$. The three principal volumes create halo samples offset horizontally, and slight differences in FP and DMO mass functions are apparent, especially at dwarf halo mass scales, $\Mtwohc < 10^{11} \msol$. The flattening at the lower mass threshold is an edge effect arising from each sample's minimum halo mass.} 
    \label{fig:Halo_Counts_effective}
\end{figure}

At a given mass, $M$, the effective number of haloes contributing to the local fit is given by the sum of weights, 
\begin{equation}\label{eqn:N_halos_effective}
    N_{\rm eff} (M) = \sum_{i = 1}^{N_{\rm haloes}} \ w_i(\mu),
\end{equation}
where $\mu = \ln(M/\!\msun)$ and the halo weight around mass $\mu$ is a Gaussian,
\begin{equation} \label{eqn:KLLR_weights}
    w_i(\mu) = \frac{1}{\sqrt{2\pi}\sigma_{\rm KLLR}} \exp \bigg \{ - \frac{(\mu_i - \mu)^2}{2\sigma_{\rm KLLR}^2} \bigg\} .
\end{equation}
Figure \ref{fig:Halo_Counts_effective} shows $N_{\rm eff}(M)$ for the FP and DMO variants of all three TNG runs at $z=0$.  At the minimum mass scale, the effective number is near or above $10^5$ while at the maximum mass it is approximately 200.

Our one modification to the \textsc{Kllr} method allows for regression on \textit{multiple} properties beyond just halo mass. We use this variant to study the scatter in $\sigmaDM$ when conditioned on secondary halo properties in addition to halo mass. We now describe the formalism of this modification.

The Gaussian kernel used for weighting all haloes is still a 1D Gaussian in $\mu$. Taking $\mathbf{X} \in \{ x_1, \ldots, x_N \}$ as a chosen subset of $N$ secondary features, the conditional expectation value, $\mathbb{E}[y \, | \, \mu, \mathbf{X}]$, of property $y$ is modelled as a multivariate linear function,

\begin{equation} \label{eqn:MKLLR_expectation_val}
    \mathbb{E}[y \, | \, \mu, \textbf{X}] = \pi(\mu) + \alpha(\mu) \mu + \boldsymbol{\beta}(\mu)\cdot\mathbf{X}\,,
\end{equation}
where the normalization, $\pi$, scaling slope with mass, $\alpha$ and vector of secondary slopes, $\boldsymbol{\beta}$, are all functions of the primary regression feature, $\mu$, and are estimated by minimizing the weighted sum of squares,

\begin{equation} \label{eqn:KLLR_Cost_function}
    \epsilon^2(\mu) = \sum_{i = 1}^{N_{\rm haloes}}w_i^2(\mu) \, \bigg(y_i - \pi(\mu) - \alpha(\mu)\mu_i - \boldsymbol{\beta}(\mu)\cdot\mathbf{X}_i\bigg)^2.
\end{equation}
The intrinsic variance in property $y$ is 
\begin{equation} \label{eqn:MKLLR_Variance}
    \sigma^2_y(\mu) = A \sum_{i = 1}^{N_{\rm haloes }} w_i(\mu)(y_i - \mathbb{E}[y \,|\, \mu_i, \textbf{X}_i])^2
\end{equation}
with 
\begin{equation} \label{eqn:Scatter_Norm}
    A = \bigg(\sum_i^{N_{\rm haloes }}w_i\bigg) \big/ \bigg[\bigg(\sum_i^{N_{\rm haloes }}w_i\bigg)^2 - \sum_i^{N_{\rm haloes }}w_i^2 \bigg]\, .
\end{equation}
The expressions above revert to equations (2) and (4) in \citet{Farahi2018BAHAMAS} when no additional properties are added. 

\subsection{Dark Matter halo properties} \label{sec:DMO_property_def}

The halo properties we study have multiple possible definitions and methods of computation.  Here we describe our approach.  A key difference with some previous work is that, for properties other than mass, we ignore all substructure and unbound particles and use only the central (or primary) subhalo of each halo.

\begin{table}
    \centering
    \begin{tabular}{l|p{0.7\linewidth}}
         \textbf{Property} & \textbf{Definition}  \\
         \hline
         $M$ & $\Mtwohc$, spherical mass ($<\Rtwohc$) of DM and baryons\\[4pt]
         $\sigmaDM$ & Isotropic DM velocity dispersion\\[1pt]
         $\ctwohc$ & NFW concentration\\[1pt]
         $\aform$ & Scale factor when $M(\aform) = 0.5 \, M(a=1)$\\[1pt]
         $\sDM$ & Density major-to-minor axis ratio\\[1pt]
         $\sDMvel$ & Velocity major-to-minor axis ratio\\[4pt]
         $\Gammadyn$ & Average mass accretion rate over one dynamical time.\\[1pt]
         $\EsDM$ & Energy content of the DM surface pressure.\\
         \hline
    \end{tabular}
    \caption{List of halo properties used in this work. Total mass is the primary independent variable used for regression of the rest. The next set of five properties form our main analysis; the last two are shown in Appendix~\ref{appx:Additional_Properties}}
    \label{tab:Halo_properties}
\end{table}

In the standard TNG analysis, haloes are first identified using a Friends of friends (FoF) percolation method, after which the \textsc{Subfind} algorithm \citep{Springel2001Subfind, Dolag2009Subfind} both uses the density field to identify subhaloes and then applies a binding energy condition to each identified subhalo to discard particles that are not bound to it. Since FoF works only in position-space, FoF haloes can contain substructure (subhaloes or particle streams) with vastly different velocities to the other substructure in the halo. We have confirmed that such ``contamination'', where velocity distributions within $\Rtwohc$ are multimodal, happens frequently for low-mass haloes lying in the infall regions of massive haloes. When left uncorrected, such multimodality strongly affects the velocity-based statistics of interest to us. Our solution is to use only the central subhalo in all of our property calculations with the exception of the halo total mass. This avoids the velocity bimodality and also provides lower-noise measurements of the DM density profile and halo shape parameters, which are all affected by substructure. The focus on only the central subhalo removes less than $10\%$ of the DM mass within $\Rtwohc$ in cluster-scale haloes, and this number drops to $3\%$ for dwarf galaxy-scale haloes.

\subsubsection{Halo Mass, $M$} \label{sec:mass_def}

We use the halo mass from the public TNG catalogs, which was computed via a spherical-overdensity condition, $M(<\Rtwohc) = \Mtwohc$, satisfying
\begin{equation} \label{eqn:M200c}
    \Mtwohc = \frac{4\pi}{3}\Rtwohc^3 \, 200\rho_c(z)\,,
\end{equation}
where $\rho_c(z)$ is the critical density of the universe at a given epoch. For every TNG run, all available types of particles are used when computing $\Mtwohc$. For the DMO variants, this involves using only DM, and for the FP variants, all of DM, stars, gas, and SMBHs are used. Note also that this computation does \textit{not} ignore substructure and unbound particles, whereas that is the case for the quantities that follow.

\subsubsection{Isotropic Dark Matter Velocity Dispersion, $\sigmaDM$} \label{sec:Sigma_DM_3D_def}

Relaxation drives the DM velocity distribution toward a Gaussian form, with a variance that scales with the depth of the gravitational potential well. We follow the early work of \citep{Yahil1977Velocity} in utilizing the angle-averaged, proper velocity dispersion of DM particles in a halo, 
\begin{equation} \label{eqn:Yahil_sigmaDM}
    \sigmaDM^2 = \frac{1}{3(N_{\rm cen} - 1)}\sum_{k = 1}^{N_{\rm cen}} \sum_{j=1}^3 \bigg(v_{j,k} - \langle{v}_{j}\rangle \bigg)^2, 
\end{equation}
where $v_{j,k}$ is the peculiar velocity (does not include the Hubble flow) for a particle $k$ within the $N_{\rm cen}$ particles of the central subhalo and index $j$ runs over principal Cartesian axes of the simulation volume. The terms $\langle{v}_{j}\rangle$ represent the mean velocity of the same set of $N_{\rm cen}$ particles.  

This isotropic definition is the average trace element of the velocity dispersion tensor, defined in section \ref{sec:shape_def}, from which we extract a minor-to-major axis ratio, $\sDMvel$.

\subsubsection{Concentration, $\ctwohc$}\label{sec:c200c_def}

We closely follow the method adopted by the \textsc{Rockstar} set of algorithms \citep{Behroozi2013Rockstar} that have been employed extensively for various simulations. Using the set of central subhalo DM particles within $\rtwoh$, we construct $50$ equal-mass radial bins between $3\epsilon < r < \Rtwohc$, where $\epsilon$ is the force softening scale of the particular run (see Table~\ref{tab:Sim_data}). We compute the densities within each bin and fit them with an NFW profile \citep{Navarro1997NFWProfile},
\begin{equation} \label{eqn:NFW_profile}
    \rho(r) = \frac{\rho_s}{\frac{r}{r_s}\big(1 + \frac{r}{r_s}\big)^2}\,,
\end{equation}
where $\rho_s$ and $r_s$ are the characteristic density and the scale radius, respectively. The fit is performed using the \texttt{COLOSSUS}\footnote{\url{https://bdiemer.bitbucket.io/colossus/}} package  \citep{Diemer2018COLOSSUS} via its least-squares minimization method. We apply a constant $15\%$ fractional uncertainty to the density computed within each bin during fitting, and the derived concentrations,
\begin{equation}\label{eqn:c200c_def}
    \ctwohc = \Rtwohc/r_s,
\end{equation}
are insensitive to this choice. Some alternative methods for computing $\ctwohc$ (or analogs of it) involve using a different functional form for the density profile \citep{Einasto1965DensityProfile}, taking the maximum of the DM circular velocity profile \citep{Gao2004Subhalos, Bose2019GalaxyHaloConnection}, and taking the ratio of halo masses enclosing different overdensities \citep{Thomas2001NFW}. Most literature has continued using the NFW concentration, and we follow this as well.

Our measurements of $\ctwohc$ for the low halo mass end --- where haloes are more poorly resolved, and can sometimes have density profiles that vary significantly from NFW --- show a handful of haloes have very high concentrations, up to $\ctwohc \sim 300$, and this is also found in several recent works \citep[\eg][]{Diemer2015Concentration, Mansfield2020AssemblyBias}. Here, we employ a simple $3\sigma$ filtering procedure across the whole halo sample in order to remove any such outliers from our analysis. Here, $\sigma$ is the local scatter measured at a given halo mass scale, $\Mtwohc$. This procedure discards $\sim 2\%$ of our haloes per run, with almost all discarded haloes residing near the lower mass limit of each run. The $\ctwohc - \Mtwohc$ relations are insensitive to this filtering step --- with one obvious exception in the local scatter --- and this is more thoroughly discussed alongside our results (section \ref{sec:c200c_Fit}).

\subsubsection{Halo formation time $\aform$} \label{sec:aform_def}

The formation time, $\aform$, is defined as the cosmological scale factor at which the halo first attained half the mass that it has at the present epoch,
\begin{equation} \label{eqn:aform}
    \Mtwohc(a = \aform) = \frac{1}{2}\Mtwohc(a = 1).
\end{equation}
In practice, we following the subhalo merger tree of the central subhalo, then identify the \textit{first} instance when the halo achieves at least half its $z=0$ mass.  We interpolate linearly in $\ln a$ and $\ln \Mtwohc$ to get $\aform$, and also discard any times in the merger tree for which the subhalo progenitor is the satellite of another halo. Such a situation is common for lower-mass haloes during flyby events where the smaller halo passes through the outskirts of a larger halo, briefly becoming its satellite. The merger trees we use were constructed using \textsc{SubLink} \citep{Rodriguez-Gomez2015SubLinkTrees}, and with only DM particles (even in the FP runs). Note that while the merger tree is based on the central subhalo's evolution, $\aform$ is derived from $\Mtwohc$ which is computed using all available particles (section \ref{sec:mass_def}).

There are also numerous other definitions for the formation time present in the literature, each with different approaches and different uses. A summary of eight such definitions based on either halo or galaxy properties is provided by \citet{Li2008FormationTimes}.

\subsubsection{Density and velocity shape parameters, $\sDM$ and $\sDMvel$}\label{sec:shape_def}

For both density and velocities, we compute the quadrupole moment tensor, $\mathcal{M}$, using the central subhalo's DM particles,
\begin{equation} \label{eqn:shapes}
    \mathcal{M}_{ij} = \frac{1}{N_{\rm cen}}\sum_{k = 1}^{N_{\rm cen}}\bigg(x_{k,i} - \langle x_i \rangle \bigg) \bigg(x_{k,j} - \langle {x}_j\rangle\bigg),
\end{equation}
where $x_{k, i}$ is the $i^{\rm th}$ Cartesian coordinate (position or peculiar velocity) of the $k^{\rm th}$ particle, and $\langle x_i \rangle$ are reference values. For the density-space shape the reference position is the most-gravitationally bound particle in the halo while for the velocity-space shape it is the mean velocity of all the DM particles in the summation. This is consistent with \textsc{Subfind}'s definitions of the position and velocity centers of haloes and subhaloes  \citep[see Section 4.2]{Springel2001Subfind}. In each case, we compute the eigenvalues of the tensor, $\mathcal{M}$, and extract the estimate --- which is exact for the case of a true ellipsoid --- of the minor-to-major axis ratio,
\begin{equation} \label{eqn:c_over_a}
s = \sqrt{\frac{\lambda_3}{\lambda_1}} , 
\end{equation} 
where $\lambda_1 \ge \lambda_2 \ge \lambda_3$ are the ranked set of  eigenvalues.  We use the notations $\sDM$ for density and $\sDMvel$ for velocity measures, and note that both lie on the unit interval, $s \in (0,1]$.

The shape parameters can be computed using many different tensors \citep{Bett2012HaloShapeEstimators}, and a thorough discussion on popular shape estimators and their associated use cases can also be found in \citet{Zemp2011ShapeEstimators}. For this work we have chosen the simplest tensor formulation.

\section{\textsc{Kllr} Statistics of TNG Halo Populations} \label{sec:Fit_Params}

We present here mass-dependent fit parameters for the scaling behavior of the five primary DM properties, listed in Table~\ref{tab:Halo_properties}, with halo mass. A key area of focus is the difference between FP and DMO parameters, and in particular, the mass-dependent features gravitationally induced into DM structure by galaxy formation processes such as gas cooling and stellar/SMBH feedback. To highlight baryon back-reaction effects, we show 
FP run parameters with colored lines and DMO one in gray-scale in each of the figures below. A second focal area is that of numerical consistency across simulation volumes with the same physics treatment. These can be gleaned by comparing sets of either colored lines or grey-scale lines for FP and DMO, respectively, in overlapping mass ranges.

All shaded bands in the figures to follow give the $68\%$ confidence intervals estimated via $100$ bootstrap realizations of the halo sample. Such bootstrap errors capture variance from sample size only, and thus should be considered \textit{lower bounds} on a ``true'' theoretical uncertainty that could contain additional components such as cosmic variance, variation in numerical methods, and (for FP runs) astrophysical model uncertainties. The dashed lines in the top panels of each figure show the tails of each distribution, via the local $95\%$ quantiles of a given property computed within mass bins of $0.2\,\, \rm dex$ width. The quantile can be exactly computed for bins with $N > 10^2$ haloes, but for sparsely populated bins the result is interpolated using the available data.

For just the slope and scatter, we apply a light boxcar $(0.25-0.5-0.25)$ smoothing to the \textsc{Kllr} measurements which are computed at $0.1\,\rm dex$ intervals. This does not affect results for $N_{\rm eff} > 10^3$ but it softens small features in the noisier, high-mass end of each run, thereby enhancing visibility of the large-scale trends that are the focus of our work. All qualitative and quantitative statements made below are unaffected by this choice.

\subsection{Dark Matter Velocity Dispersion, \texorpdfstring{$\sigmaDM$}{sigmaDM}} \label{sec:Sigma_DM_3D_Fit}
 
\begin{figure}
    \centering
    \includegraphics[width = \columnwidth]{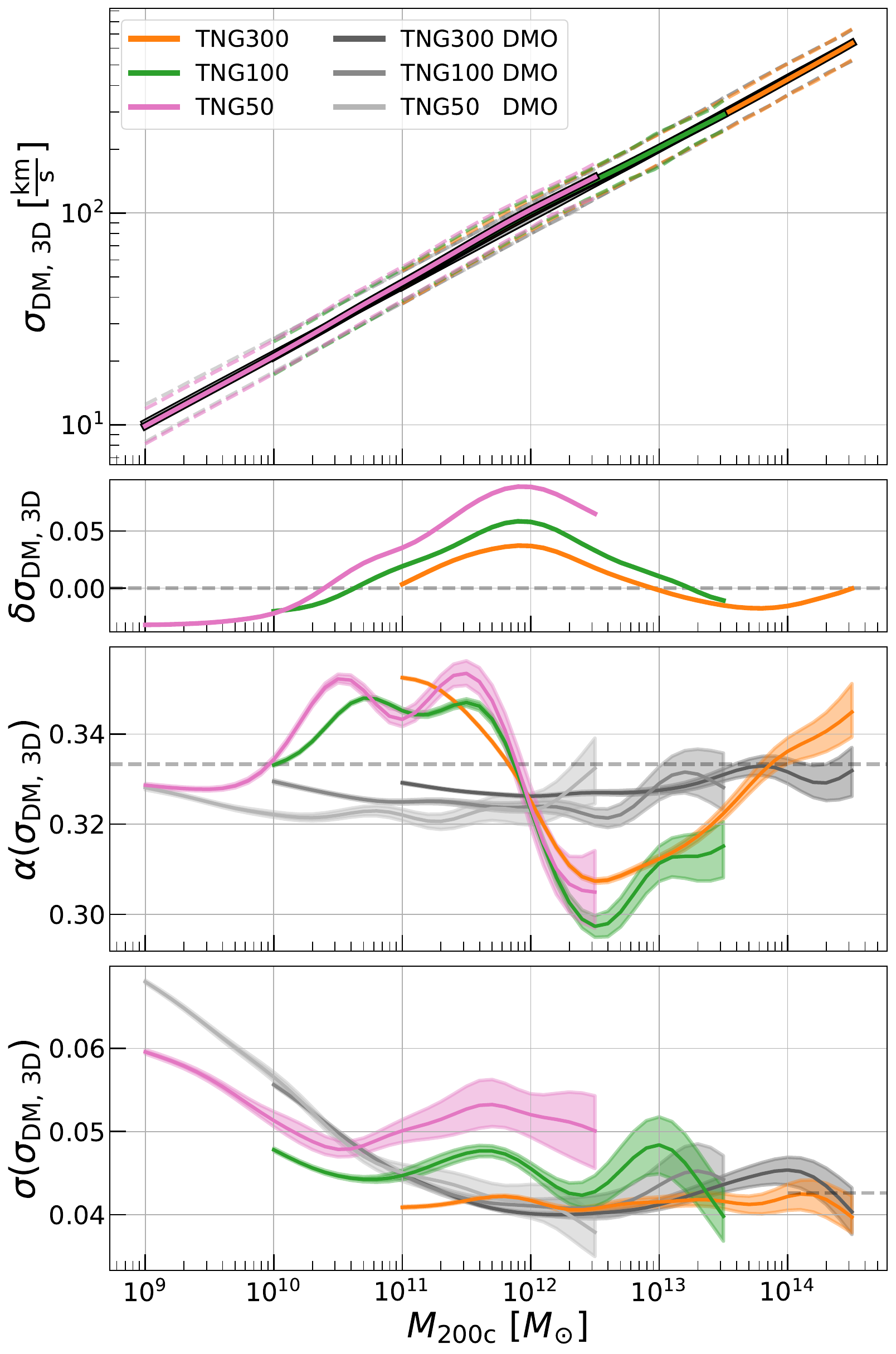}
    \caption{The $z=0$ \textsc{Kllr} scaling parameters of isotropic DM velocity dispersion, $\sigmaDM$, with halo mass, $\Mtwohc$, for the DMO (grey) and FP (colors) runs. The four panels, from top to bottom, show (i) normalization, (ii) fractional difference, $\delta X = \ln(X^{\rm FP}/X^{\rm DMO})$, (iii) slope, and; (iv) population intrinsic scatter in natural log (so a fractional scatter). The dashed lines in the top panel show the $\pm 2\sigma$ bounds of $\sigmaDM$ in mass bins of $0.2\, \rm dex$, and the bands in the bottom two panels show the $1\sigma$ bounds estimated from $100$ bootstrap realizations. The fractional difference is maximized at MW masses, with amplification from resolution. 
    These differences are also clearly seen in the slope --- the DMO run is close to the virial expectation, $\alpha = 1/3$ (dashed gray line), while the FP run significantly deviates from this. The intrinsic population scatter (bottom panel), which is expressed in natural log terms, is amplified by resolution in the FP runs and is consistent with the \citet{Evrard2008VirialScaling} estimate of $4.2\%$ (gray dashed line) at high masses.}
    \label{fig:sigma_DM_3D_summary}
\end{figure}

Under strict self-similarity, the $\sigmaDM - \Mtwohc$ scaling relation should show a slope $\alpha = 1/3$ if one assumes the virial theorem \citep{Zwicky1937Virial}.  A meta-analysis using large samples of massive haloes ($\Mtwohc > 10^{14} \msol$) extracted from DMO and non-radiative simulations verifies these expectations, finding $\alpha = 0.3361 \pm 0.0026$ \citep{Evrard2008VirialScaling}. In the lower-middle panel of Figure~\ref{fig:sigma_DM_3D_summary}, we extend this measurement of $\alpha$ to lower halo masses, $\Mtwohc = 10^{9} \msol$, and use \textsc{Kllr} to extract mass- or resolution-sensitive trends.  At galaxy cluster scales ($\Mtwohc > 10^{14} \msol$), the DMO and FP slopes are consistent with $1/3$, but deviations occur at lower halo masses.   

For the DMO halo population, the slopes are essentialy constant but show mild resolution dependence, with $\Delta \alpha \approx 0.005$ between pairs of runs. These slopes also kick up toward $0.33$ at the lowest resolved mass scales in each volume, and we suspect this is an unphysical feature induced by resolution limits. On the other hand, the prediction of $\alpha < 1/3$ for masses below $10^{14} \msol$ is consistent with predictions from an analytic derivation by \citet{Okoli2016Concentration} and is likely a physical result.

The DMO population's intrinsic scatter, shown in the lowest panel, is also fairly insensitive to numerical resolution.  At high halo mass, both the DMO and FP treatments produce populations with a scatter\footnote{The intrinsic population scatter is shown in natural log terms (not dex) and thus is a fractional scatter.} that agrees with the value of $0.0426 \pm 0.015$ (dashed line) found by \citet{Evrard2008VirialScaling}. Starting at $10^{12} \msol$, the DMO population scatter rises toward dwarf galaxy mass scales, and this increase is closely-related to an analagous increase found in $\ctwohc$ with origins discussed further in Section \ref{sec:c200c_Fit}.

In contrast to the more regular behavior of the DMO models, the FP runs create a halo population whose DM features have statistically significant, mass-dependent deviations.  The colored lines in the upper-middle panel of Figure~\ref{fig:sigma_DM_3D_summary} display a mass-localized increase in $\sigmaDM$ for FP haloes at $\Mtwohc \sim 10^{12} \msol$.  We will refer to this feature as a ``wiggle'' because it corresponds to increasing and decreasing slopes at lower and higher mass-scales, respectively. The FP runs can be up to $\approx 8\%$ higher than the DMO run counterparts for this scaling relation. For galaxy group and cluster scales ($>10^{13} \msol$) the FP and DMO runs are in agreement at the few percent level, a finding consistent with previous works focused on high mass haloes \citep{Lau2010BaryonDisspiation, Munari2013VelDispBaryons, Armitage2018C-EagleVelDisp}.

The FP wiggle feature has clear resolution dependence; its amplitude increases with resolution. This trend likely arises from the dependence of the star-formation efficiency (SFE) on resolution in the TNG model \citep{Pillepich2018FirstGalaxies}. At low redshifts, the stellar mass of a MW-sized galaxy is the dominant interior mass component of its halo. Thus, increasing stellar mass can deepen a halo's gravitational potential well, increasing the equilibrium velocity dispersion. Given SFEs and stellar masses are larger in the higher resolution runs \citep{Pillepich2018FirstGalaxies}, one should expect the velocity dispersions to be higher as well, as is evident.  We show below that this picture is also consistent with features in the $\ctwohc$ scaling relation.

For haloes of MW mass and lower, the scatter in the FP runs behaves in both mass- and resolution-dependent ways. It increases with numerical resolution for $\Mtwohc < 10^{12.5} \msol$, and exhibits regimes with both larger and smaller scatter than the DMO runs.
The scatter for galaxy cluster-scale haloes was previously shown to be unaffected by the inclusion of FP physics \citep{Lau2010BaryonDisspiation, Munari2013VelDispBaryons, Armitage2018C-EagleVelDisp}, and we find this to be the case for halo masses of $\Mtwohc > 10^{13} \msol$.

Spectroscopic measurements of galaxies only probe line-of-sight velocities, so we also examined population statistics using each velocity dispersion component of a halo separately.  By construction, the resulting $\sigmaDMOneD - \Mtwohc$ relation matches the normalization and slope behavior of the 3D average, but the anisotropic shape of the velocity tensor enhances the scatter. Thus, $\sigmaDMOneD$ has a scatter of $\approx 6\%$ for most halo masses which then increases to $\approx 8\%$ for $\Mtwohc > 10^{13} \msol$, where recent mergers enhance the average velocity anisotropy (lower $\sDMvel$) of the population. This quantity is part of our public data release and can be explored in more detail by the interested reader.

In section \ref{sec:Impact_Feedback}, we redo this analysis for the TNG model variant runs and find that the location, in halo mass, of the FP wiggle feature is sensitive to the parameters of the SMBH feedback model.

\subsection{NFW Concentration, \texorpdfstring{$\ctwohc$}{c200c}} \label{sec:c200c_Fit}

\begin{figure}
    \centering
    \includegraphics[width = \columnwidth]{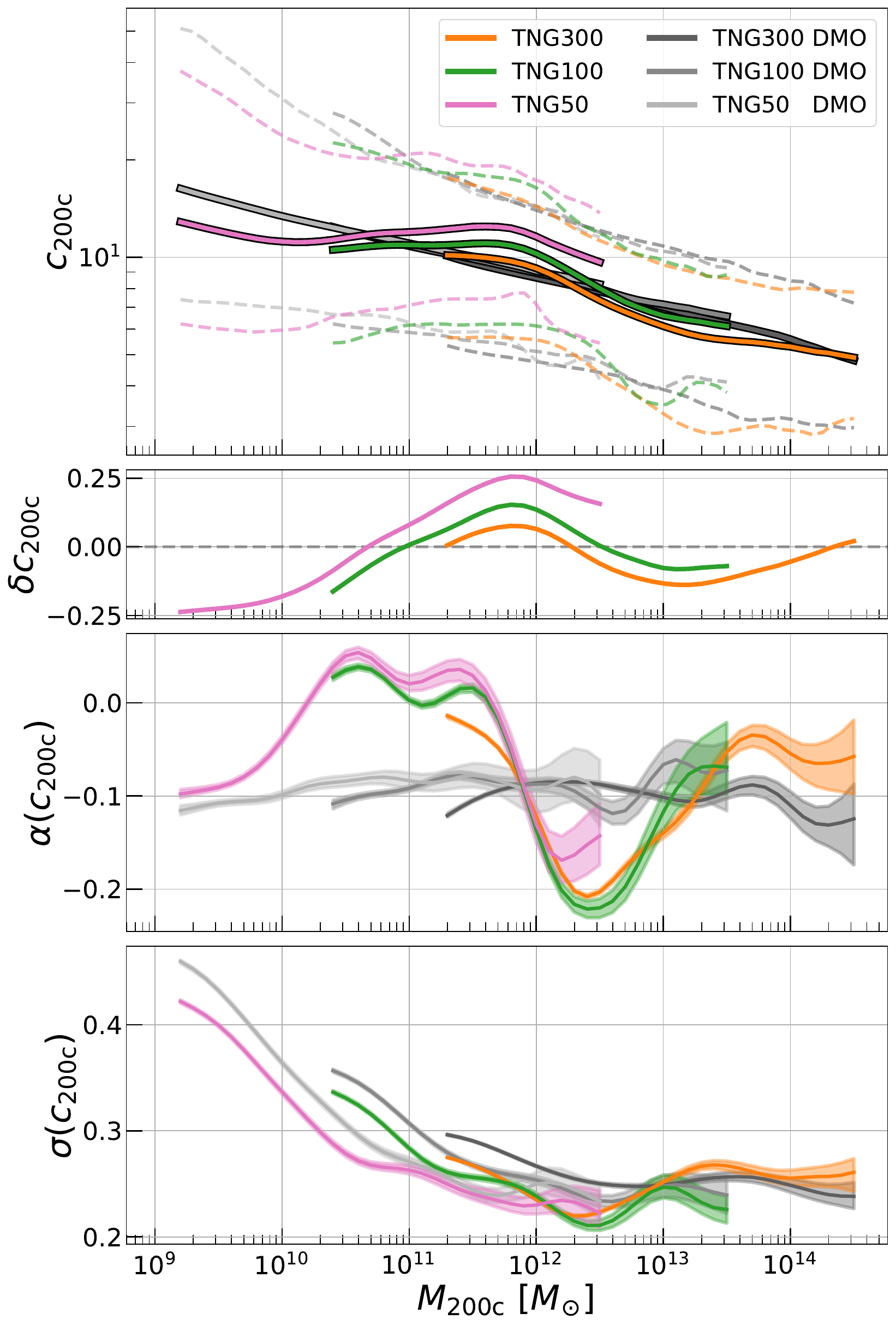}
    \caption{The $z=0$ \textsc{Kllr} scaling parameters of DM NFW concentration, $\ctwohc$, 
    using the same style conventions as Fig.~\ref{fig:sigma_DM_3D_summary}. The FP runs show clear differences from their DMO counterparts, with a bump near $\Mtwohc \approx 10^{12} \msol$ that increases with resolution,  and deficits at both higher and lower mass scales as induced by different feedback mechanisms (section \ref{sec:Impact_Feedback}). The fractional differences can be up to $25\%$ as shown in the upper middle panel. Due to the more stringent resolution requirement for computing $\ctwohc$, the halo mass threshold here is a factor $\sim 2$ larger than that listed in Table~\ref{tab:Sim_data}.}
    \label{fig:c200c_summary}
\end{figure}

For haloes in dynamical equilibrium, the concentration and dark matter velocity dispersion will be coupled through the gravitational potential. Indeed, we show in \S\ref{sec:explain_scatter} below that these two properties are strongly correlated at fixed halo mass, especially at MW-scales and lower which tend to be dynamically older and therefore closer to equilibrium.  

The top two panels of Figure~\ref{fig:c200c_summary} show the mean $\ctwohc - \Mtwohc$ relation. Once again, features from the inclusion of baryons are evident; the FP mean relations are either higher or lower than the DMO ones depending on the halo mass-scale. The inclusion of baryons undergoing dissipation tends to increase the DM density in the halo core through adiabatic contraction  \citep[\eg][]{Blumenthal1986AdiabaticContraction, Gnedin2004AdiabaticContraction}, and the strength of this effect can depend on the baryon assembly history \citep{Abadi2010ShapesBaryons, Tissera2010BaryonImprints, Pedrosa2010BaryonImprints, Artale2019BaryonImprints, ForouharMoreno2021BaryonsConcentrationEagle}. Since the efficiency of dissipative processes peaks near MW masses, the concentration in the FP runs is boosted at that scale. However, competing feedback processes, from SN at lower masses and SMBHs at higher masses, disperse baryons out of the core and into the outer halo, thus decreasing the DM concentration at these mass scales. Above $\sim 10^{14}\msol$, the shift in mean concentration is very small. 

This effect has been shown previously for cluster-scale haloes \citep{Duffy2010BaryonDmProfileDensity, Cui2016NiftyBaryonsHaloProperties}, MW-scale haloes \citep{Pedrosa2010BaryonImprints, Artale2019BaryonImprints, ForouharMoreno2021BaryonsConcentrationEagle} and for dwarf galaxy-scale haloes \citep{Wetzel2016DwarfGals, Wheeler2019FIREdwarfs}. The impact of SMBHs underlies the strongest feature we see --- the steeper drop in $\ctwohc$ (compared to DMO runs) at $\Mtwohc = 10^{12} \msol$, which is the average halo mass-scale where the SMBHs kinetic feedback mode is activated and thus proceeds to drive the DM distribution into a less-concentrated profile.

Previously, \citet[see their Figure 5]{Lovell2018ConcentrationBaryonImprints} and \citet[see their Figure 6]{Beltz-Mohrmann2021BaryonImpactTNG} used the TNG100/300 runs to study baryonic imprints on \textit{proxies} of $\ctwohc$. Both works show mass-dependent, baryon-induced deviations consistent with those shown in Figure~\ref{fig:c200c_summary}, but \citet{Beltz-Mohrmann2021BaryonImpactTNG} still concluded that the property relation was insensitive to baryons. This is because their methodology --- which was to fit a simple power-law to the averages computed in mass bins, and then compare log-linear slopes across runs --- is a zeroth-order comparison and is thus insensitive to the ``wiggle''-type deviations detailed in this work.

Next, as was the case with $\sigmaDM$, there is clear resolution dependence for $\ctwohc$ in the FP runs, with increasing resolution leading to more concentrated haloes at MW-scales. Once again, the driving influence is likely the impact of resolution on SFEs. Having more stars in the core deepens a halo's gravitational potential and leads to more concentrated haloes in the higher resolution FP runs. Note also that for $10^{10.5} \msol < \Mtwohc < 10^{11.5} \msol$ the mean concentration actually 
\textit{increases} with $\Mtwohc$, and this behavior is more prominent at higher redshifts (see section \ref{appx:RedEvo_c200c}).

The mean relations in the DMO runs are insensitive to resolution and also consistent with recent simulation-based models \citep[\eg][]{Ludlow2016Concentration, Child2018ConcentratioMassRelation, Diemer2019concentrations, Ishiyama2020UchuuConcentration},
though the level of agreement does vary across halo mass.
In general, our mean concentration at dwarf galaxy-scales ($\Mtwohc \sim 10^{9} \msol$) is within $5\%$ of the predictions from the models mentioned above, but is up to $20\%$ higher at galaxy cluster-scales ($\Mtwohc > 10^{13.5} \msol$), with the exact deviations depending on the model. The deviation and its increase with halo mass arise from differences in computing $\ctwohc$; the aforementioned works use \textit{all} DM particles within $r < \Rtwohc$, whereas we only use those from the central subhalo.

The mass-dependent slopes in the DMO runs, shown in the lower middle panel of Figure~ \ref{fig:c200c_summary}, are remarkably close to a constant value of $\alpha = -0.1$ and are consistent with previous DMO analyses which found values ranging from $\alpha \in [-0.08, -0.12]$ \citep[\eg][]{Bullock2001Concentrations, Neto2007Concentration, Duffy2008Concentration, Bhattacharya2013Concentrations, Klypin2016Concentrations, Child2018ConcentratioMassRelation}.   The FP slopes vary more dramatically, and also show a zero-crossing corresponding to a local-maxima in the mean relation near $\Mtwohc = 5 \times 10^{11}\msol$. For cluster-scales, the slope tends toward $\alpha = -0.05$ and is larger than the DMO counterpart --- both features are consistent with previous findings \citep{Henson2017BaryonEffectsBM}. Work by \citet{Ragagnin2019HaloConcentration} found $\alpha \approx -0.12$ for a suite of hydrodynamics simulations, but they fit a single power law to about four decades in halo mass.

The population intrinsic scatter, shown in the bottom panel of Figure~\ref{fig:c200c_summary}, is relatively constant at $\sim 25\%$ for $\Mtwohc > 10^{12} \msol$ with the FP runs showing slightly lower scatter. This weak or nonexistent trend of the scatter over this mass range agrees with previous predictions \citep{Neto2007Concentration, Diemer2015Concentration}, though our scatter estimate of $\sim 25\%$ is lower compared to previous estimates of $32 - 36\%$ \citep[\eg][]{Bullock2001Concentrations, Wechsler2002Concentrations, Neto2007Concentration, Duffy2008Concentration, Bhattacharya2013Concentrations, Diemer2015Concentration}. However, these works find that a re-analysis using just \textit{relaxed} haloes obtains estimates of scatter around $20 - 25\%$. Similarly, if we forego the $3\sigma$ outlier filtering step in our analysis then our scatter increases to $30\%$, in better agreement with fiducial estimates using all available haloes. Note that the filtering steps changes the mean relation and the slopes by less than $1\%$.

The scatter also has a significant, mass-dependent increase below $\Mtwohc < 10^{11} \msol$, caused by the tail to high $\ctwohc$. This tail is almost entirely accounted for by low-mass haloes that reside near more massive haloes and have had their DM phase-space altered via strong tidal interactions with these massive neighbors \citep[see Figure 6 in][]{Mansfield2020AssemblyBias}. We have confirmed that removing such haloes subdues the increase in the scatter at low masses seen in $\ctwohc$ \textit{and} in $\sigmaDM$.

\subsection{Halo formation time, \texorpdfstring{$\aform$}{aform}} \label{sec:aform_Fit}

\begin{figure}
    \centering
    \includegraphics[width = \columnwidth]{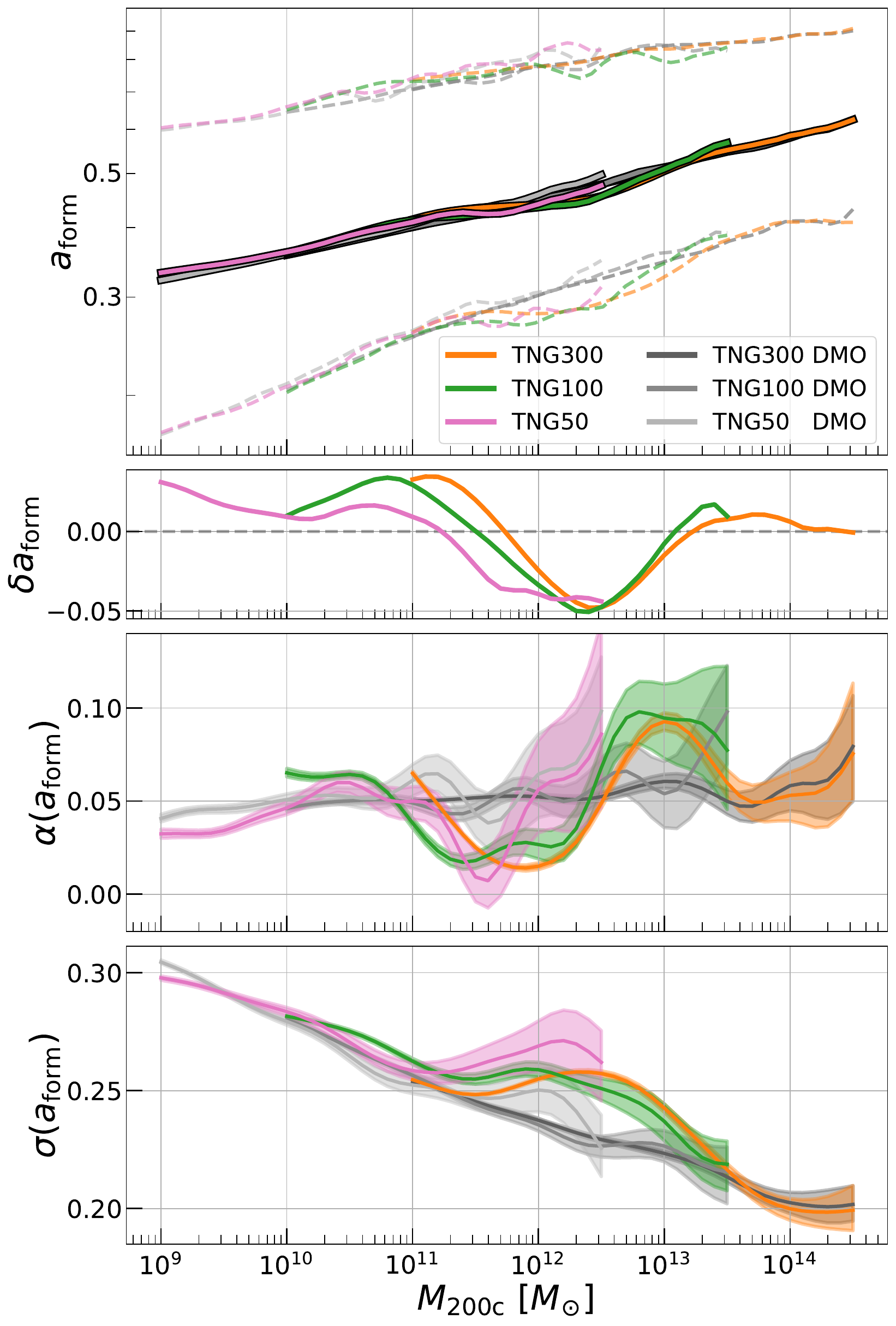}
    \caption{The $z = 0$ \textsc{Kllr} scaling parameters of halo formation time, $\aform$,
    using the same style conventions as Fig.~\ref{fig:sigma_DM_3D_summary}. MW-scale haloes form $\approx 0.4 \gyr$ earlier in the FP runs, and haloes below this mass scale form $\approx 0.2 \gyr$ later. The scatter in the FP runs is also amplified around MW-scale haloes.}
    \label{fig:a_form_summary}
\end{figure}

The mean formation epoch, $\aform$, increases with mass, consistent with the canonical picture of bottom-up structure formation where larger haloes form later via mergers of smaller haloes. The top panel of Figure~\ref{fig:a_form_summary} shows that the FP treatment shifts formation times of the MW mass haloes to be earlier than that of the DMO treatment, with a maximum difference of $\approx 0.4 \gyr$.  At lower masses, the formation times are mildly delayed. The mean relations agree with previous results \citep{Neto2007Concentration, Li2008FormationTimes, Bose2019GalaxyHaloConnection}, who used the same definition and collectively estimate a range $\aform \in [0.4, 0.6]$ for a corresponding halo mass range of $M \in [10^{11}, 10^{14}] \msol$.

The wiggle feature in the FP treatment is seen once again in the slopes.  Increasing numerical resolution tends to shift the feature to somewhat lower mass-scales, but the effect is not large.  The slopes of the DMO treatment are fairly insensitive to resolution and lie within $\alpha \in [0.04, 0.06]$ over the six decades in mass shown. This range is consistent with the estimate of $\alpha = 0.046$ from \citet{Neto2007Concentration}.

The fractional scatter in both DMO and FP runs declines with increasing halo mass. The scatter in the FP treatment has a mass-localized increase between $10^{11} \msol < \Mtwohc < 10^{13} \msol$, indicating that the inclusion of baryons causes haloes in this mass range to form over a wider range of epochs. The dashed lines in the top panel of Figure~\ref{fig:a_form_summary}, which show the $95\%$-ile range, also display a noticeable shift to earlier epochs in this mass range.

\subsection{Density and velocity shape parameters, \texorpdfstring{$\sDM$}{sDM} and \texorpdfstring{$\sDMvel$}{sDMvel}}

\begin{figure*}
    \centering
    \includegraphics[width = \columnwidth]{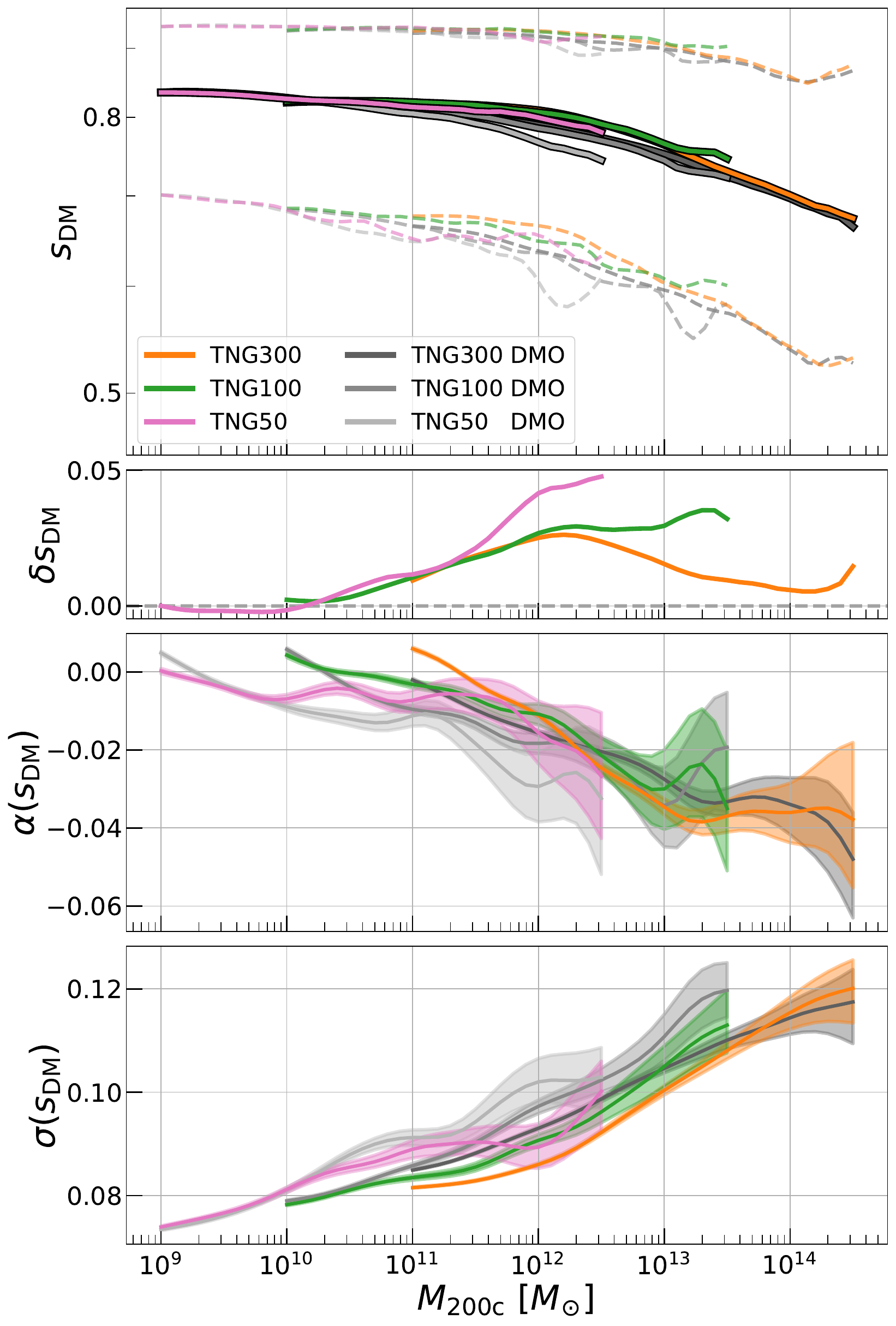}
    \includegraphics[width = \columnwidth]{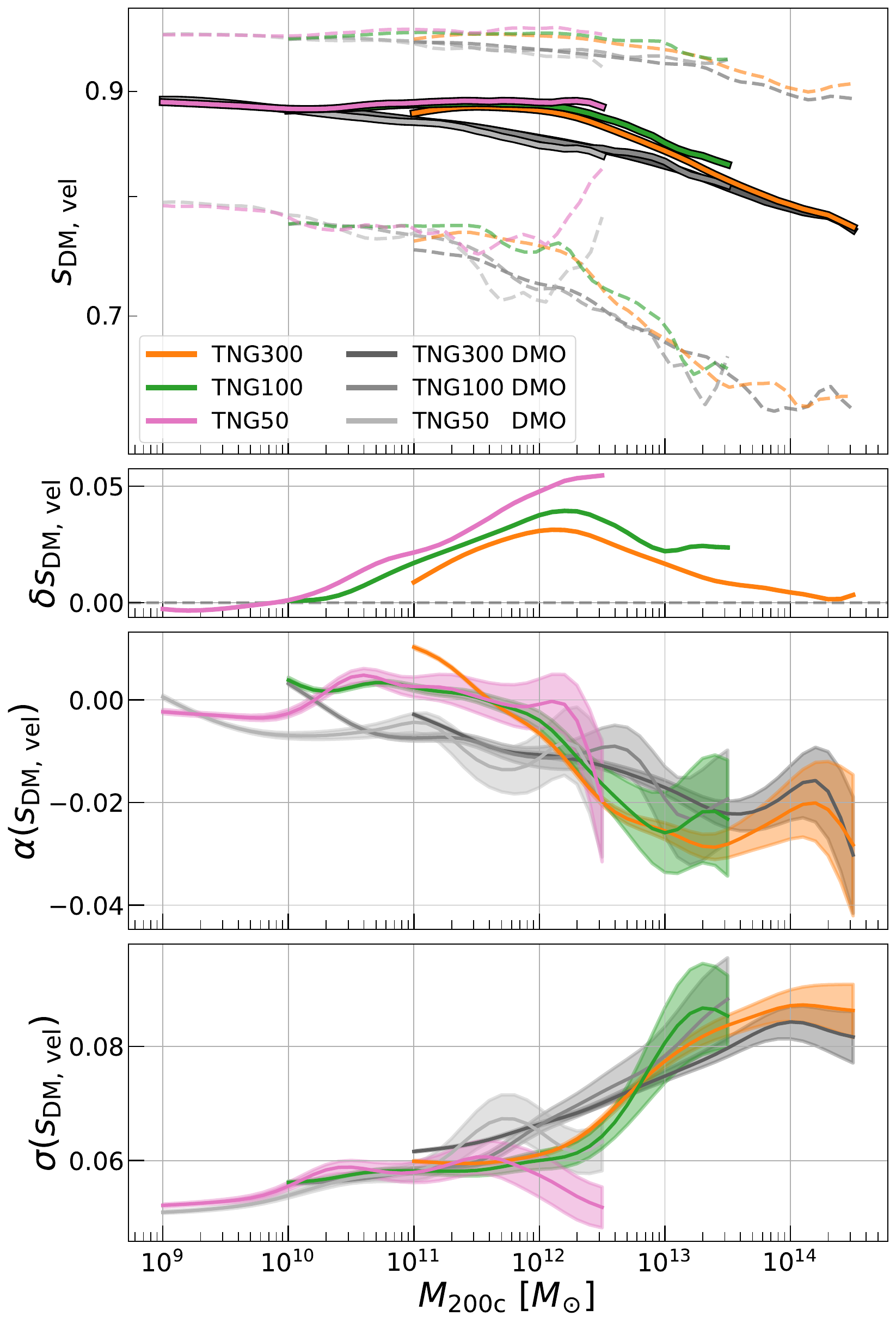}
    \caption{The $z = 0$ \textsc{Kllr} scaling parameters of the halo density-space shape, $\sDM$ (left), and velocity-space shape, $\sDMvel$ (right), defined as the ellipsoidal minor-to-major axis. We use the same style conventions as Fig.~\ref{fig:sigma_DM_3D_summary}. The addition of baryons results in a pronounced rounding of haloes (higher $\sDM$ and $\sDMvel$ values) near the MW mass scale of $\Mtwohc \approx 10^{12} \msol$, the mass scale of peak SFE, and this boost is stronger for $\sDMvel$ as the velocity ellipsoid responds to the gravitational potential, which is rounder than the underlying density field due to the Laplacian connecting the two. Shapes for dwarf galaxy- and cluster-scale haloes are insensitive to galaxy formation when computed within $r < \Rtwohc$. The scatter increases monotonically with mass in both treatments, with FP showing a slight diminution near the MW mass scale. In the FP runs the mean $\sDMvel$ is essentially constant, $\sDMvel = 0.9$, below $\Mtwohc < 10^{12} \msol$.
    }
    \label{fig:shapes_summary}
\end{figure*}

Figure~\ref{fig:shapes_summary} shows the scaling behaviors of the density- and velocity-space shapes in the left and right columns, respectively.  Recall that $\sDM$ and $\sDMvel$ are minor-to-major axis ratios found by diagonalizing the particle density and velocity tensors, defined in \S\ref{sec:shape_def}. Given that more massive haloes form at later times, and have thus had less time to dynamically relax, the massive halo population tends toward a more ellipsoidal mean shape (lower $\sDM$ and $\sDMvel$), and a larger fractional scatter with more significant tails extending to lower shape values.

Early baryonic simulations with radiative cooling found that the inclusion of dissipation via the gas made galaxy-scale haloes rounder \citep{Evrard1994TwoFluidGalaxyFormation} and this result has been replicated with more detailed baryonic physics treatments  \citep[\eg][]{Kazantzidis2004HaloShapes, Abadi2010ShapesBaryons}.
For both DM shape parameters, we find the mean relations in the FP runs are higher than those of the DMO runs, and this is most pronounced near the MW mass scale, where the fractional deviations can be up to $5\%$ with some mild resolution dependence. The shapes at dwarf galaxy-scales and cluster-scales are less sensitive to galaxy formation processes.

The strength of the FP deviations depends somewhat on the radial scale within which the shape parameters are computed. In general, the impact of baryons is most prominent within the inner halo $(\sim 0.2 \Rtwohc)$, so when computing the shape within $\Rtwohc$, we expect less prominent, but still noticeable, differences between the DMO and FP runs \citep{Kazantzidis2004HaloShapes}. However, when gas cooling is the only relevant baryonic process, the impact on shapes computed within $\Rtwohc$ may still be significant \citep{Debattista2008HaloShapesCooling, Abadi2010ShapesBaryons}. This is relevant for haloes below the MW mass-scale where gas cooling is the dominant baryonic mechanism (see section \ref{sec:Impact_Feedback}), and indeed in Figure~\ref{fig:shapes_summary} the deviations between FP and DMO begin at $\Mtwohc \approx 10^{10.5} \msol$.
Recent work on $\sDM$ has shown that haloes tend to be rounder due to the inclusion of galaxy formation physics \citep[\eg][]{Knebe2010Shapes, Lau2011ShapesOfGas, Bryan2013BaryonImpactOnShapes, Tenneti2015ShapesHydroVsDMO, Butsky2016NihaoHaloShapes, Cui2016NiftyBaryonsHaloProperties, Henson2017BaryonEffectsBM, Chua2019ShapeIllustrisBaryons, Chua2021TNGShapes}, while varying the galaxy formation components, such as stellar or AGN feedback, do not influence this effect much \citep[\eg][]{Cui2016NiftyBaryonsHaloProperties, Velliscig2015EagleOwlsHaloShapes, Chua2021TNGShapes}. The estimated scatter is insensitive to the inclusion of baryons and this is likely a product of the large aperture, $r < \Rtwohc$, that we compute the shapes within. 

Haloes are also rounder in velocity-space than density-space, as shown previously for cluster-scale haloes using both DMO simulations \citep{Kasun2005ShapesAlignments} and observations \citep{Wojtak2013HaloShapesSDSS}. Notably, the FP run mean relation asymptotes to a nearly constant value of $\sDMvel \approx 0.9$ over a mass range $\Mtwohc = 10^{9} - 10^{12} \msol$. Such behavior is not exhibited by the DMO runs. We note that the results for both shape parameters are also specific to the tensor we use to obtain the shapes (section \ref{sec:shape_def}). Differences in the utilized tensor may explain the fact that even in the DMO runs, our $\sDM$ values are higher than those of some previous works \citep{Bonamigo2015MilleniumHaloShapes, Vega-Ferrero2017MdplDmoHaloShapes}.

The slopes in $\sDM - \Mtwohc$ (left column, middle panel, figure \ref{fig:shapes_summary}) show weak sensitivity to the inclusion of baryons, in agreement with previous results for cluster-scale haloes \citep{Bryan2013BaryonImpactOnShapes, Henson2017BaryonEffectsBM}. \citet{Allgood2006Shapes} also quote a power-law slope of $\alpha = -0.05$ for haloes of $\Mtwohc \approx 10^{12} - 10^{14} \msol$, which is steeper than our findings. However, they compute shapes within $0.3 R_{\rm vir}$ which is a significantly smaller aperture than that chosen for this work.

\subsection{Observable Consequence: Concentration vs. Central Galaxy Stellar Mass} \label{sec:Observations}

\begin{figure}
    \centering
    \includegraphics[width = \columnwidth]{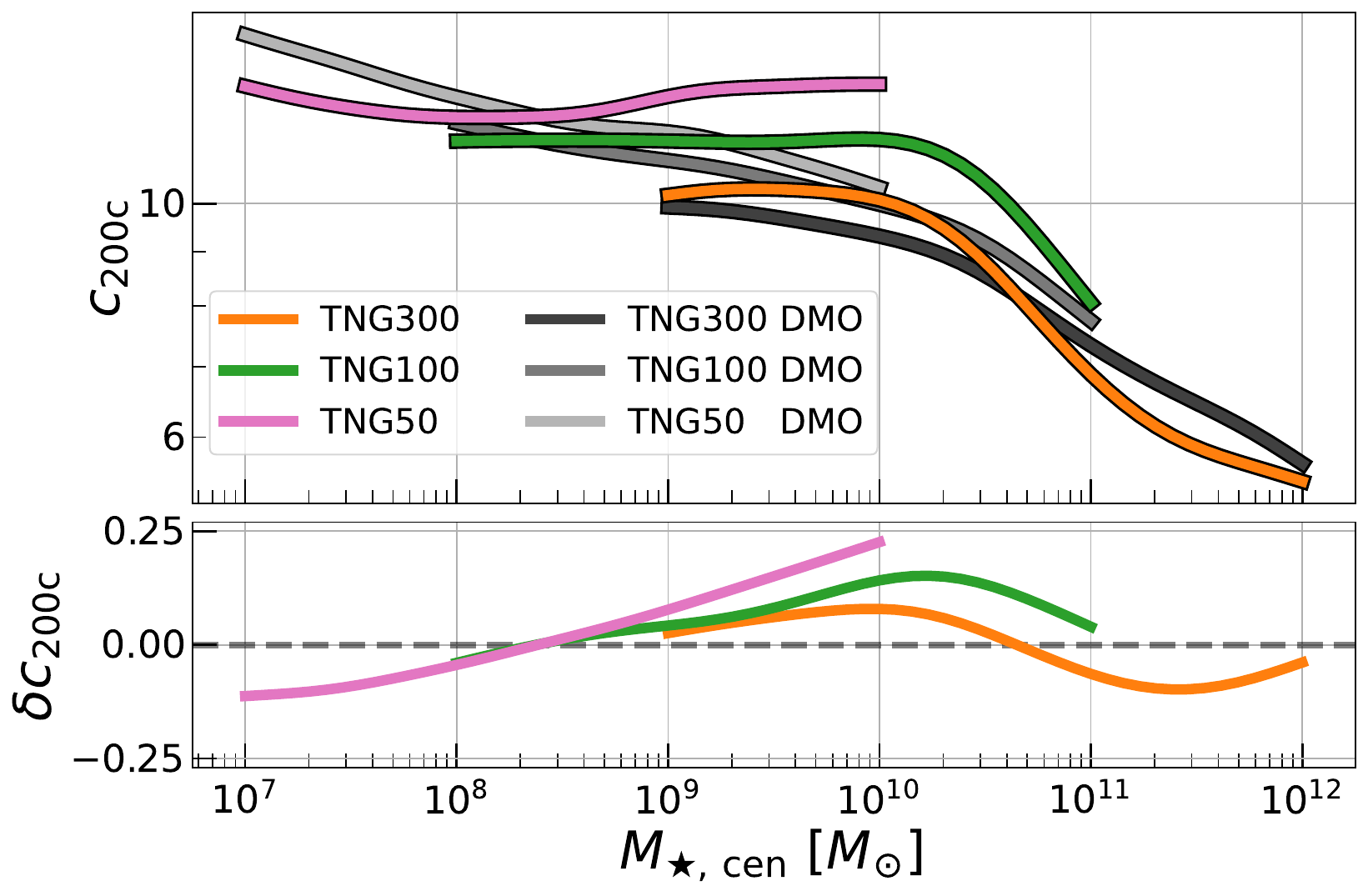}
    \caption{The $z=0$ \textsc{Kllr} scaling relation of halo concentration, $\ctwohc$, as a function of its central galaxy stellar mass, $\mstarcen$. Both quantities are observationally accessible.  The mean $\ctwohc$ in the FP runs is nearly constant below $\mstarcen = 10^{10} \msol$, in contrast with the DMO runs where it shows a more gradual decrease. The bottom panel shows the fractional difference in log-mean amplitude, $\delta \ctwohc = \ln (\ctwohc^{\rm FP} / \ctwohc^{\rm DMO})$. The DMO run haloes are assigned $\mstarcen$ by matching each with its counterpart in the FP run. See text for details.}
    \label{fig:MStar_Observable_c200c_Rockstar}
\end{figure}

The scaling relations described above share a common trait: the FP treatment produces a halo population with properties that have mass-dependent ``wiggles'', centered at the MW mass of $\approx 10^{12} \msol$, that are not present in the DMO runs. The mass-localization of the wiggles is a consequence of cooling and feedback processes that we discuss in section \ref{sec:Impact_Feedback}. In this section, we address the observability of these mass-localized features by replacing the total mass of a halo with a key observable proxy, the stellar mass of the halo's central galaxy/subhalo, $\Mstarcen$.

Figure \ref{fig:MStar_Observable_c200c_Rockstar} shows the $\ctwohc - \Mstarcen$ scaling relation at $z=0$ in the FP and DMO models. The DMO haloes are assigned a stellar mass by matching each one to its corresponding halo in the FP run, and we utilize the public TNG matching catalog for this task. This matching was done using the \textsc{SubLink} merger tree weighting algorithm \citep[see equation 1]{Rodriguez-Gomez2015SubLinkTrees}, which was run on only DM particles. The algorithm was originally designed to match subhaloes between different redshifts of the same simulation, and was adopted to match subhaloes between \textit{different} simulations at the same redshift. Note that the matching across runs is possible only because the FP and DMO runs both start with the same initial conditions. Around $95 - 98\%$ of haloes in our FP sample are assigned a match in the corresponding DMO sample.

In the DMO treatment, the $\ctwohc - \Mstarcen$ relation has an approximately power-law form (Figure \ref{fig:MStar_Observable_c200c_Rockstar}), with $\alpha \approx -0.075$.  The FP relations show significant curvature and deviations from this form within the mass range $10^{10} \msol < \Mstarcen < 10^{12} \msol$. The fractional amplitudes of these wiggles can be up to $25\%$; there is a clear resolution dependence in this estimate, but the wiggle feature itself exists regardless of resolution.

While we have explored a wide mass range in Figure \ref{fig:MStar_Observable_c200c_Rockstar}, the existence of the FP features can also be observationally verified for just high-mass haloes ($\Mtwohc > 10^{13} \msol$). At this regime, additional, more reliable mass-proxies become available, such as X-ray temperature and luminosity. For example, the \texttt{eROSITA} all-sky X-ray survey \citep{Merloni2012eROSITAScienceBook} will detect $N \approx 10^{5}$ haloes of galaxy group-scales and above, and of a median redshift, $z = 0.35$ \citep{Pillepich2012eRositaForecast}. Next, the halo concentration is frequently inferred from profiles of the weak gravitational lensing signal \citep[\eg][]{Mandelbaum2008ConcentrationMeasurement, Shan2012ConcentrationObs, Covone2014ConcentrationObs, Umetsu2014ConcentrationObs, Merten2015ConcentrationObs}, which are sensitive to the projected mass distribution of the halo. Observational measurements from \citet{Shin2021MassGalaxyProfilesDES} show that the galaxy number density profile can also serve as a good proxy. Stacking haloes to boost the lensing signal is also possible as we are only interested in the mean relation. Measurements of galaxy-galaxy lensing are also a potential avenue; \citet{Zacharegkas2021GGLensingDES} used a $\ctwohc-\Mtwohc$ relation alongside a halo occupation distribution (HOD) framework to fit small-scale galaxy-galaxy lensing measurements ($r > 70 \kpc$) and constrain cosmology. However, one could fix the cosmology and HOD parameters, and then employ the same methodology to constrain the $\ctwohc - \Mtwohc$ relation instead.

\subsection{Impact of Feedback} \label{sec:Impact_Feedback}

\begin{figure*}
    \centering
    \includegraphics[width = 2\columnwidth]{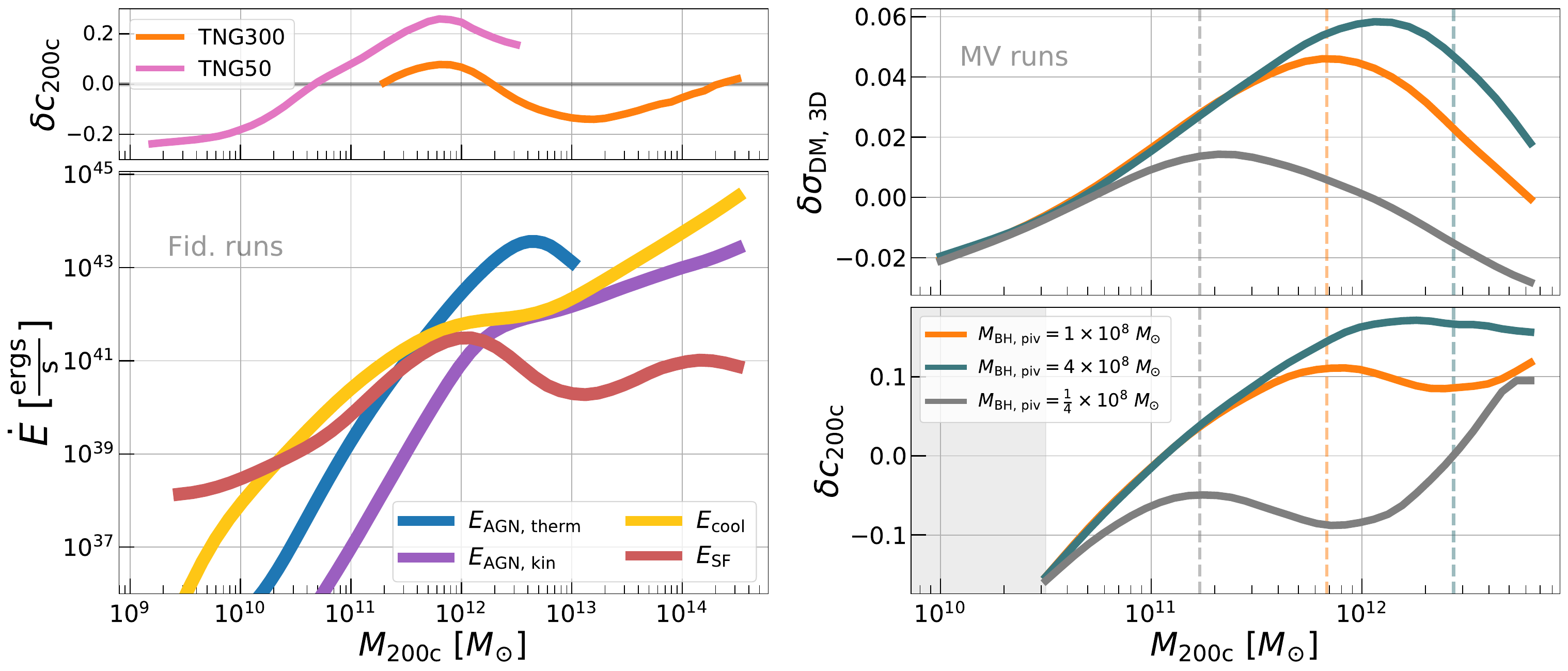}
    \caption{\textbf{Left:} The top panel shows the fractional difference in halo concentration, $\delta \ctwohc = \ln(c_{\rm 200c}^{\rm FP}/c_{\rm 200c}^{\rm DMO})$, for the fiducial TNG50/300 runs. The bottom panel shows the average \textit{instantaneous} energy rate of different baryonic processes as a function of halo mass. We combine all three TNG FP runs to produce one curve per process. The halo mass-scales where different processes take over coincide with the variations between the FP and DMO runs shown in the top panel. For gas cooling losses, we show the absolute value of the energy rate. \textbf{Right:} The fractional difference $\delta X = \ln (X / X^{\rm DMO})$ for three model variant (MV) runs --- one that uses the fiducial FP model (orange) and two that vary $\MBHpiv$. Results are shown for velocity dispersion (top) and halo concentration (bottom). Dashed vertical lines, a factor of $4$ apart in $\Mtwohc$, approximately capture shifts in the mass-localized features driven by changes in $\MBHpiv$. 
    }
    \label{fig:Feedback_Variation}
\end{figure*}

In general, all the mass-localized features in the DM scaling relations under FP treatment are sourced by the gravitational coupling between the dark matter and the baryons of multiple phases. The phases are governed primarily  by: (i) Gas cooling; (ii) Stellar feedback; (iii) SMBH thermal feedback (quasar mode), and (iv) SMBH kinetic feedback (radio mode). The latter three inject energy into the gas phase of the halo while the first is an energy loss process.

Each process has an associated instantaneous energy rate, $\dot{E}$, which we compute for each halo by summing the individual contributions of particles/cells associated with the central subhalo. The cooling rates per gas cell are provided in the public TNG data release, while the stellar and SMBH feedback energies were computed as detailed in \citet{Weinberger2018SMBHsIllustrisTNG}. 

The bottom left panel of Figure~\ref{fig:Feedback_Variation} shows the average $\dot{E}$ as a function of halo mass, and it is clear that each process has a halo mass range where it dominates the others by at least an order of magnitude. Note that we have combined the halo samples of TNG50/100/300 to produce one curve per process. There \textit{are} some resolution differences between the runs, but these are subdominant compared to the order-of-magnitude differences between the processes. The top left panel of Figure~\ref{fig:Feedback_Variation} shows the fractional differences between the $\ctwohc - \Mtwohc$ relations of the FP and DMO runs for TNG50 and TNG300. As different physical processes compete to dominate the baryon plasma's evolution, the core gravitational potential of the halo will be forced accordingly.  The DM halo population properties, as a result, are imprinted with features localized at halo masses where the dominant physical mechanism changes.

Now, we systematically detail the dominant feedback mechanism at a given halo mass, and, as an example, the impact made on $\ctwohc$ as was previously noted in Section \ref{fig:c200c_summary}. First, stellar feedback is most effective for the smallest haloes, and blows out material from their shallow potential. This decreases $\ctwohc$. Second, gas cooling dominates for about a decade in mass which, through adiabatic contraction, leads to more concentrated haloes. Third, the SMBH thermal feedback activates, and stifles gas cooling while also dispersing matter out of the halo core. Both effects lead to less concentrated haloes. Finally, the SMBH population shifts dominantly into its kinetic mode, and gas cooling dominates once again, leading towards more concentrated haloes. 

We then explicitly test the dependence of the MW-scale features in the scaling relations on the TNG SMBH feedback model. The right column of Figure~\ref{fig:Feedback_Variation} displays the fractional deviations between the mean relations of three model variant FP runs with respect to an associated DMO run, and this is shown as a function of halo mass for the velocity dispersion (top panel) and concentration (bottom panel). The MW-scale features shifts up/down in halo mass as the black hole pivot mass, $M_{\rm BH,\,piv}$, as defined in equation \eqref{eqn:BH_Accretion_Mode}, is increased/decreased by factors of four; recall that $M_{\rm BH,\,piv}$ influences the average SMBH mass (and thus average halo mass) at which the transition from quasar mode to radio mode feedback occurs. The response of the wiggle feature's location to changes in $M_{\rm BH,\,piv}$, with consistent behavior shown across multiple separate halo properties, provides strong evidence that the feature is connected to the SMBH feedback mechanisms.

\begin{figure}
    \centering
    \includegraphics[width = \columnwidth]{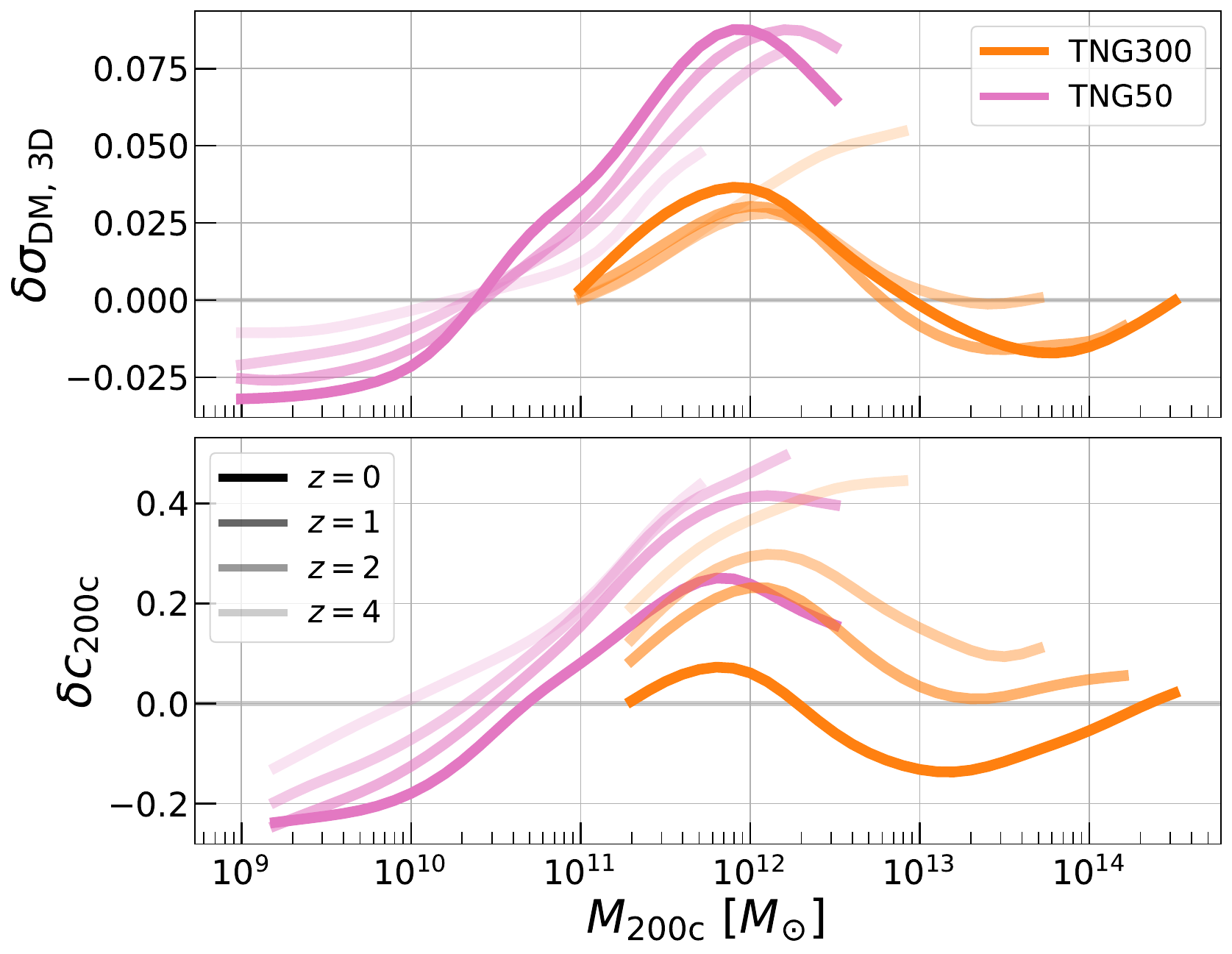}
    \caption{The fractional difference $\delta X = \ln (X^{\rm FP} / X^{\rm DMO})$ for the velocity dispersion (top) and halo concentration (bottom), for a redshift range $0 < z < 4$ which corresponds to a time interval $\Delta t \approx 12.5\gyr$. The baryon-induced wiggles exists all redshifts, though the specific temporal evolution of their shape and amplitude depends on the halo property --- differences in concentration decrease with time and those in velocity dispersion increase with time.}
    \label{fig:Wiggle_Redshift_evol}
\end{figure}

Note that Figure~\ref{fig:Feedback_Variation} only shows the \textit{instantaneous} energy rates, whereas the features at $z = 0$ in the FP runs are generated over some period of time. In Figure~\ref{fig:Wiggle_Redshift_evol}, we also show that the FP run features persist at all redshifts, though their temporal evolution has property-specific behaviors --- deviations in the velocity dispersion grow with time, whereas those in concentration become more \textit{suppressed}. Previous findings also show the difference between FP and DMO for $\ctwohc$ are more pronounced at higher redshifts \citep{Duffy2010BaryonDmProfileDensity, Cui2016NiftyBaryonsHaloProperties}. Note that since baryonic process have a non-trivial radial dependence, the redshift behavior of the differences could also depend on the aperture that a given property is computed within.

The dominance of stellar feedback at low halo masses, and of AGN feedback at high halo masses picks out the MW mass-scale as the transition regime; the special status of this mass-scale is consistent with previous theoretical and observational results on the stellar mass--halo mass relation \citep{Behroozi2010SMHMRelation, Moster2010SMHMRelation, Yang2012GalaxyHaloConnection, Wang2013SMHMRelation, Behroozi2013StarFormationHistory, Reddick2013GalaxyHaloConnection, Moster2013SFR, Birrer2014GalaxyEvolution, Lu2015SFR, Rodriguez-Puebla2017GalaxyEvolution,  Shankar2017GalaxyProperties, Kravtsov2018SMHMRelation, Behroozi2019UniverseMachine}. See \citet{Wechsler2018GalaxyHaloConnection} for a review on the methods used to get these constraints.

\section{Analysis of Variance}
\label{sec:explain_scatter}

So far we have detailed the mean relations for five DM halo properties and the scatter about these relations. However, the interactions between these properties, i.e. their correlation/covariance, are also important for analyses that include two or more of these properties. In this section, we extend on existing literature by studying these correlations over a wide mass range, but in a halo mass-dependent way. First, we present the mass-dependent correlation matrix (Section \ref{sec:covariance}) estimated using \textsc{Kllr}. Next, we explore which properties control the scatter about the $\sigmaDM - \Mtwohc$ relation using both \textsc{Kllr} and an interpretable machine learning approach, with a specific focus on comparing their predictions. 

\subsection{Property Covariance}\label{sec:covariance}

\begin{figure*}
    \centering
    \includegraphics[width = 1.6\columnwidth]{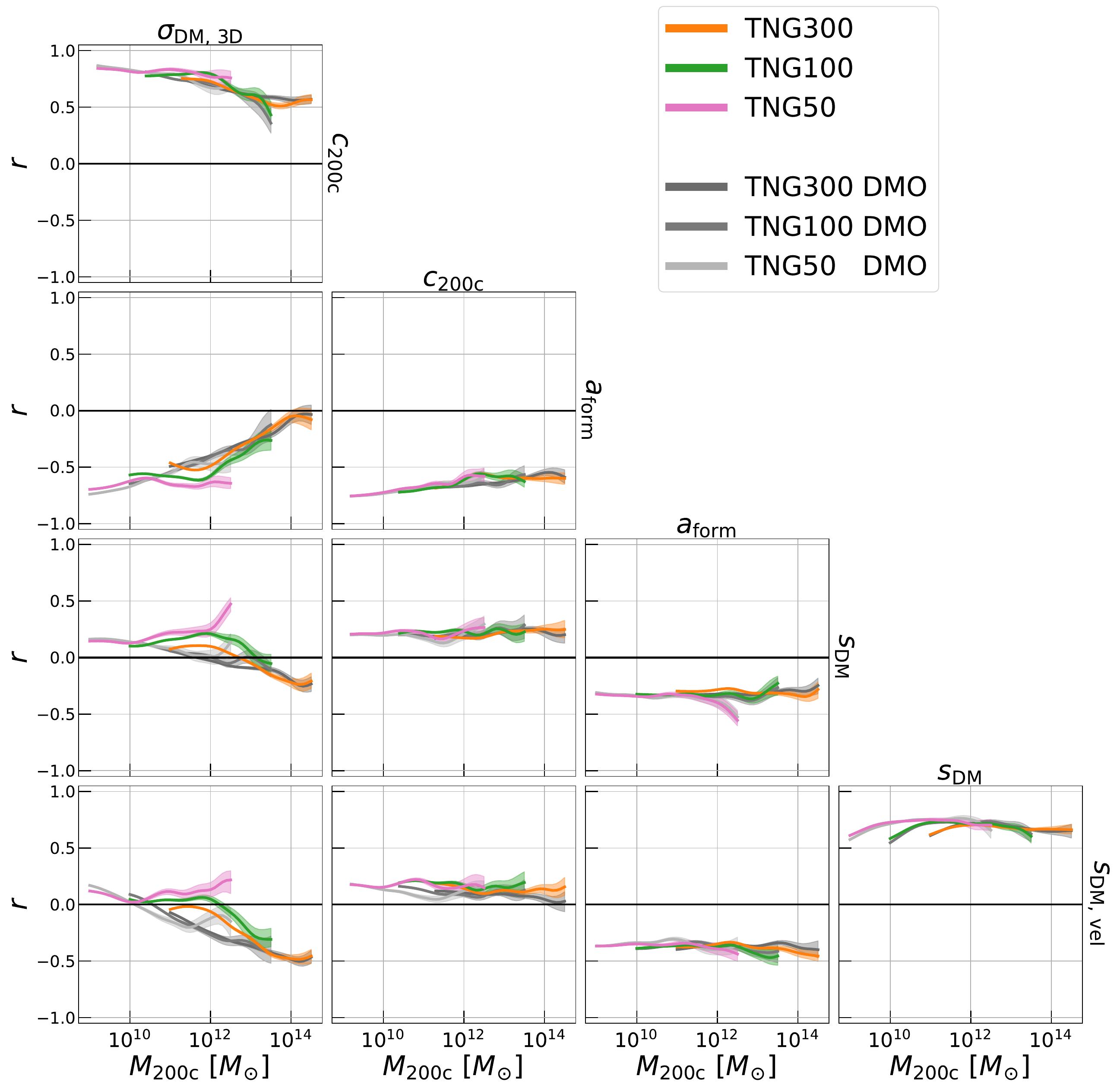}
    \caption{The $z=0$ \textsc{Kllr} correlation coefficients between pairs of DM halo properties given by the respective column and row title.  The black solid line in each panel indicates the uncorrelated limit ($r = 0$). Note that any panel with $\ctwohc$ uses a slightly different halo sample than the rest as we perform a $3\sigma$ outlier rejection for $\ctwohc$. This rejection results in 1-2\% differences in the correlations with $\ctwohc$.}
    \label{fig:Correlation_Matrix}
\end{figure*}

Here we show the $z=0$ \textsc{Kllr} correlation matrix, $r_{ij}$, for the ten unique pairings formed from the five key properties discussed in the previous section.  Standard deviations, $\sigma_i$, whose squares are the diagonal elements of the property covariance matrix, can be found in the bottom panels of Figures~\ref{fig:sigma_DM_3D_summary}, \ref{fig:c200c_summary}, \ref{fig:a_form_summary}, and \ref{fig:shapes_summary}. The off-diagonal elements of the symmetric covariance matrix, $\mathcal{C}_{ij}$, can be formed as $\mathcal{C}_{ij} = r_{ij} \sigma_i \sigma_j$. Note that \textsc{Kllr} provides a \textit{mass-dependent} estimate of the covariance matrix, $\mathcal{C}_{ij}(\mu) = r_{ij}(\mu) \sigma_i(\mu) \sigma_j(\mu)$, where the argument, $\mu = \ln(\Mtwohc/\msol)$, uses the halo mass as the scale parameter\footnote{Halo mass is a conventional choice for a halo scale parameter, but other dark matter properties, such as peak circular velocity, could also serve this role \citep{Voit2005Clusters}.  When thought of as ``labels'' applied to a population, any property that correlates strongly with halo mass could serve as the scale parameter for the \textsc{Kllr} analysis. An analytic population model based on multivariate, log-normal, mass-conditioned statistics provides the means to convert the mass-conditioned quantities presented here to quantities conditioned on other halo properties \citep{Evrard2014MultiProperty}.}.

We have already shown in \S \ref{sec:Fit_Params} that the $z=0$ scatter, $\sigma_i(\mu)$, is sensitive to halo mass and sometimes to baryonic physics as well. For MW masses, the mid-point of our mass range, the properties inversely ranked by fractional scatter are: concentration and formation epoch ($\sim\!0.25$), density-space ($\sim\!0.090$) and velocity-space ($\sim\!0.065$) shapes, and 3D DM velocity dispersion ($\sim\!0.045$).  

Figure \ref{fig:Correlation_Matrix} shows the correlation matrix of the ten unique pairings. Outside of elements paired with $\sigmaDM$, the correlation structure is independent of baryonic physics treatments. In the four-property subspace of concentration, formation epoch and the two shape measures, $r_{ij}(\mu)$ is independent of baryonic physics and nearly independent of halo mass as well. For example, the $\ctwohc-\aform$ anti-correlation gently declines from $r = -0.75$ to $r = -0.60$ over nearly $6$ decades in halo mass.

The two shape parameters show consistent, strong correlations ($r \simeq 0.7$) between each other and moderate correlation with formation epoch as well. The correlations involving $\sigmaDM$ all show clear mass-dependence and also a baryon physics sensitivity that is most pronounced at the MW mass-scale. At dwarf galaxy-scales, both concentration and formation epoch correlate strongly with $\sigmaDM$. At cluster-scales, the concentration continues to be strongly correlated, whereas $\aform$ becomes essentially uncorrelated. Also important for the $\sigmaDM$ of group and cluster-scale haloes is the velocity shape parameter, $\sDMvel$. The next section examines in detail the 3D velocity dispersion scatter and its dependence on the other properties. In Appendix \ref{appx:Correlation_Matrix}, we provide an extended version of the correlation matrix, adding mass accretion rate and surface pressure energy to the core five properties and providing results at both $z = 0$ and $z = 1$.

\subsection{Explaining the scatter in \texorpdfstring{$\sigmaDM$}{sigmaDM} with \textsc{Kllr} and machine learning}\label{sec:sigdmScatter}

\begin{figure}
    \centering
    \includegraphics[width = \columnwidth]{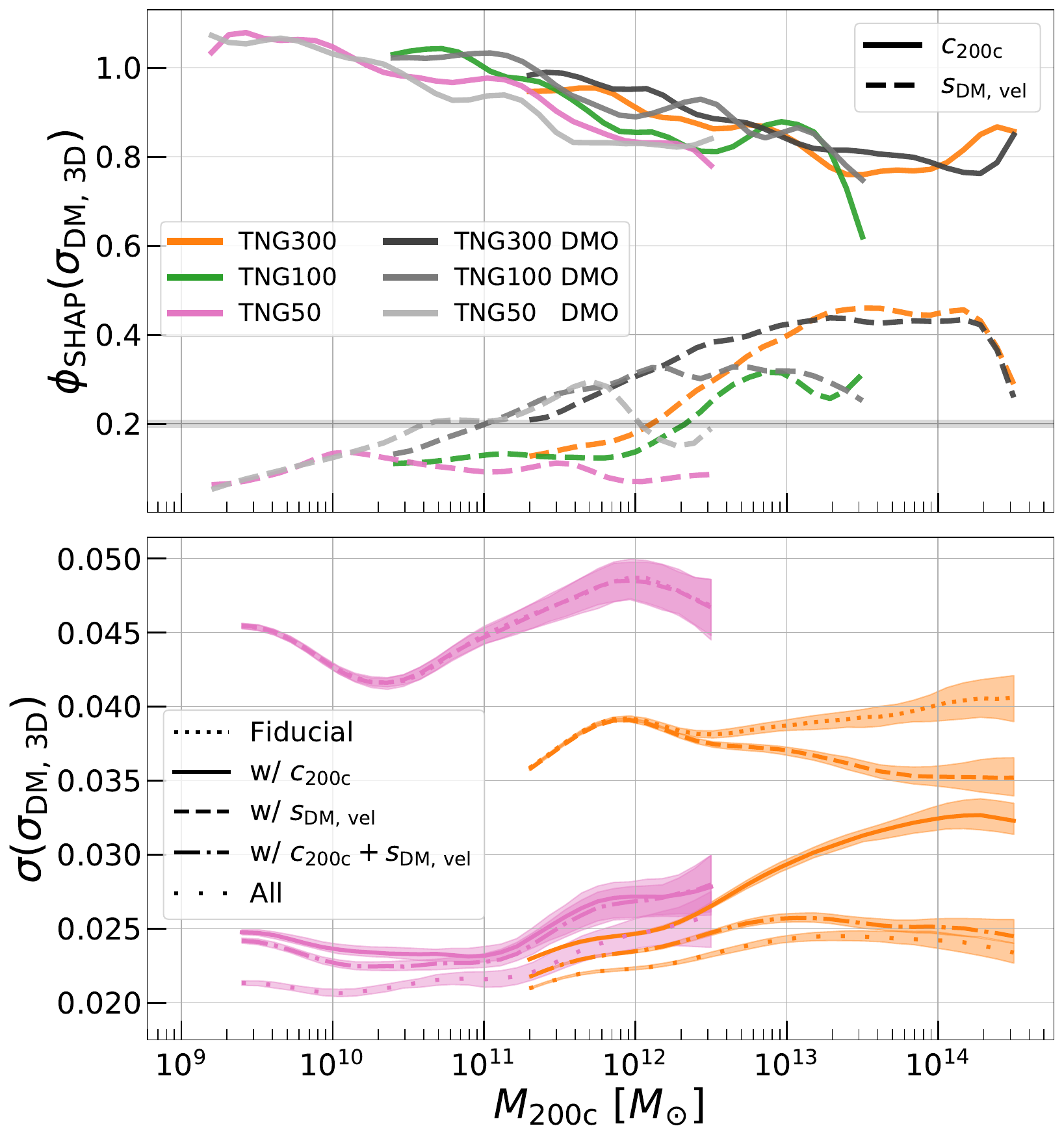}
    \caption{\textbf{Top:} The importance of halo properties in explaining the scatter about the $\sigmaDM - \Mtwohc$ relation, as determined by the $\pSHAP$ metric. The $\ctwohc$ is the most important property across all mass scales, with $\sDMvel$ becoming more relevant at cluster-scales. All other properties have contributions below $\pSHAP = 0.2$ (gray solid line), and are omitted for brevity. \textbf{Bottom:} The scatter in $\sigmaDM$ from regressing on just $\Mtwohc$ (fiducial, dotted line), or when using additional halo properties as well. We show only the TNG50/300 FP runs for brevity, and all other runs share the same behavior. The $\ctwohc$ reduces scatter at all masses, and $\sDMvel$ at high halo mass. Using $\ctwohc$ and $\sDMvel$ simultaneously reduces the instrinsic scatter down to $\approx 2.5\%$. Adding the other four halo properties does not improve on this by much.}
    \label{fig:SHAP_analysis}
\end{figure}

Next we explore which halo properties control the intrinsic scatter about the $\sigmaDM - \Mtwohc$ relation. This can be approached using \textsc{Kllr} and the correlation matrix framework, where properties that are correlated with $\sigmaDM$ --- i.e. $r_{ij} \neq 0$ --- can explain some of the intrinsic scatter. An alternative approach is to employ an interpretable machine learning (ML) framework to answer the same question. Here, we employ both techniques and find interesting similarities and differences between the two.

The \textsc{Kllr} method (section \ref{sec:KLLR}) captures potentially non-linear behaviors in halo properties as a function of one variable, in our case the log halo mass, $\mu$. The modified version described in this work can also minimize the variance in a given property by expanding the regression to include multiple additional properties beyond halo mass, as described in equation~\eqref{eqn:MKLLR_expectation_val}. In either case, the interactions between properties are completely captured by the property covariances when the population statistics are truly gaussian. However, simulated samples of massive haloes can exhibit significant non-Gaussian features --- such as those seen in both satellite galaxy counts and the central galaxy stellar mass across multiple simulations \citep{Anbajagane2020StellarStatistics} --- and indeed, all the properties described in this work also show a halo mass-dependent skewness, $\gamma(\mu) \neq 0$, that have values within the interval $\gamma(\mu) \in [-1, 1]$. The two slight exceptions to this are the velocity shape and the mass accretion rate, both of which show $|\gamma| \approx 1.5$ over a large part of the halo mass range. We have defined $\gamma$ as,
\begin{equation} \label{eqn:Skew_definition}
    \gamma(\mu) = \frac{E[(x - \Bar{x}(\mu))^3]}{E[(x - \Bar{x}(\mu))^2]^{3/2}}
\end{equation}
where $E[(x - \Bar{x}(\mu))^2]$ is more familiarly known as the variance $\sigma^2$ and $\Bar{x}(\mu)$ is the mean at log halo mass $\mu$. Here, $x$ is the log variance (except for mass accretion rate where it is the linear variable), and so the skewness is defined with respect to a Gaussian. The skewness is included in the scaling relation quantities that we make public and can be explored in more detail for each property/run/redshift.

Given our property statistics show $\gamma \neq 0$ and are thus not exactly Gaussian, machine learning (ML) presents an interesting, alternative approach to \textsc{Kllr} for predicting the value of a property, $y$, given a vector of properties $\textbf{X}$. Below, we briefly describe the two tools used for our ML analysis.

\subsubsection{Interpretable Machine Learning pipeline}
\label{sec:Interpretable_ML}

\textbf{Our first tool} is \textsc{XGBoost} \citep{Chen2016XGBoost}, a class of random forest decision tree regressors with a gradient boosting functionality that enables faster convergence to the minima of a chosen optimization scheme. Our target variable is the \textit{normalized residual} of property $y$, defined by its \textsc{Kllr} fit parameters,
\begin{equation} \label{eqn:normalized_residuals}
    \widetilde{\delta y}_{i} \equiv  = 
    \frac{y_{i} - \mathbb{E}[y \,|\, \mu_i]} {\sigma_y(\mu_i)},
\end{equation}
where $\mu_i$ and $y_i$ are the log mass and property value of halo $i$, while $\mathbb{E}[y \,|\, \mu_i]$ and $\sigma_y(\mu_i)$ are the local mean and standard deviation from regressing on just $\mu$, as defined in equations \eqref{eqn:MKLLR_expectation_val} and \eqref{eqn:MKLLR_Variance}. In this way, a normalized residual captures how many standard deviations away a halo is from the mean expectation at that halo mass. 

For each halo, the \textsc{XGBoost} model predicts $\widetilde{\delta y} _i$ given the set of features, $\textbf{X}$.  We do not include the halo mass in the feature set as the normalized residuals are constructed such that the mass-dependence is already removed. Our results are also insensitive to the choice of the \textsc{XGBoost} model hyper-parameters, and we perform our analysis using the default \textsc{XGBoost} settings.

However, machine learning models are commonly known to be black boxes; while we obtain a prediction from a model, it is unclear how the input features influenced the final outcome. Recent work in astrophysics has alleviated part of the black-box nature through interpretable ML frameworks \citep{Ntampaka2019XrayClustersML}, and such approaches are critical to enable more extensive use of ML models in the field \citep{Ntampaka2020MLinCosmoDecadal}.

\textbf{Our second tool}, the Shapley Additive Explanations \citep[\textsc{Shap}, ][]{Lundberg2020SHAP}, is designed to provide this interpretability. \textsc{Shap} is an implementation of a game theoretic metric, called the Shapley value \citep{Shapley1953SHAP}, which fairly rewards players in a game based on their contributions to the final outcome. In regression problems, each input feature acts as a ``player'' and the predicted value of the target feature is the final outcome. Thus, in the context of this work, the \textsc{Shap} method can pick out which input halo properties have the most power, according to an ML model, in predicting a target halo property. The utility of this framework was also showcased recently by \citet{Machado2020GasShapesSHAP} who used both tools to study the gaseous component of cluster-scale haloes and its relationship with DM halo properties.

For exact calculations of the Shapley contributions, the computational complexity scales exponentially with the number of input features, but the \textsc{Shap} implementation introduces an approximate method for tree-based regressors, such as \textsc{XGBoost}, that is computationally efficient and accurate. \textsc{Shap} also provides two estimation methods: (i) \texttt{interventional} and (ii) \texttt{tree path-dependent}. These differ in how they handle correlated input features. The \texttt{tree path-dependent} method is better suited for undertstanding the training data, i.e. halo properties, as opposed to the trained \textsc{XGBoost} model and is our choice for this work. A more detailed discussion on these two methods is provided in \citet{Chen2020TrueToDataModel}. 

The output of the \textsc{Shap} algorithm is a contribution measure, $\nu_{\rm SHAP}(x)$, which can be thought of as the difference between the ML prediction for $y$ given a set of input features, $\mathbf{A}$ (which includes feature $x$), and the prediction for $y$ given $\mathbf{A}$ but with feature $x$ removed. This quantifies by how much $x$ changes the target prediction. In this work, we instead show a meta-parameter, $\phi_{\rm SHAP}$ --- the mean, fractional contribution of a property $x$ to the target prediction,
\begin{equation} \label{eqn:nu_shap}
    \phi_{\rm SHAP} (x) = \frac{1}{N_{\rm haloes}}\sum_i^{N_{\rm haloes}} \frac{\nu_{{\rm SHAP}, \,i}(x)}{\widetilde{\delta y}_i}\,,
\end{equation}
where $N_{\rm haloes}$ is the number of haloes and $\widetilde{\delta} y_i$ is the normalized residual, equation~\eqref{eqn:normalized_residuals}, of the target property for halo $i$.  A value of $\pSHAP(x) = 1$ indicates that property $x$ can perfectly predict the target $y$ whereas $\pSHAP(x) = 0$ indicates that $x$ contains no useful information about $y$. The \textsc{Shap} contributions, $\nu_{\rm SHAP}$, are at most the same order as the target property (i.e. $\widetilde{\delta} y_i$), and so the individual ratios in equation \eqref{eqn:nu_shap} are at most $\mathcal{O}(1)$. An instance of the exact equality $\widetilde{\delta y}_i = 0$ does not occur for any of the $1.5$ million haloes used in this analysis. We compute $\pSHAP$ in halo mass bins of width $ \sigma_{\rm KLLR} = 0.2\,\, \rm dex$. From efficiency considerations, we also randomly downsample and use at most $2000$ haloes in each bin. Our results are insensitive to these choices.

\subsubsection{Results} \label{sec:KLLRvsMLResults}

Our main results are presented in the top panel of Figure~\ref{fig:SHAP_analysis}, where we show the contribution of $\ctwohc$ and $\sDMvel$ in explaining the normalized residuals in $\sigmaDM$. We do not show the contributions of the other, less-important parameters for brevity, and instead denote an upper limit for their individual contributions which is at $\pSHAP = 0.2$. Note that our analysis in this section also uses the mass accretion rate and DM surface pressure energy (both described in Appendix \ref{appx:Additional_Properties}).

For all haloes, the scatter is best explained by $\ctwohc$ which, like $\sigmaDM$, is also sensitive to the halo potential. Previous work provides first-principles insight into this connection \citep{Okoli2016Concentration}. The importance of $\ctwohc$ diminishes slightly for more massive haloes, and this is accompanied by an increasing importance of $\sDMvel$. The latter is expected, as the definition of $\sigmaDM$ in equation \eqref{eqn:Yahil_sigmaDM} assumes isotropy in velocity-space, but we have shown $\sDMvel \neq 1$ (Figure~\ref{fig:shapes_summary}), indicating that most haloes are elliptical in this space. Redoing our analysis of variance but on the one-dimensional $\sigma_{\rm DM,\, 1D}$ instead shows that $\sDMvel$ is no longer an important parameter (figure not displayed), whereas $\ctwohc$ still remains a significant one.

The bottom panel of Figure~\ref{fig:SHAP_analysis} shows the \textsc{Kllr} scatter for the TNG50/300 FP run from regressing on different sets of halo properties, and these results follow from the information in the correlation matrix (Figure~\ref{fig:Correlation_Matrix}). Using $\ctwohc$ reduces scatter over all mass-scales, with additional help from $\sDMvel$ at cluster-scales. Using all available DM halo properties does not give a significant improvement over using just $\ctwohc$ and $\sDMvel$. While $\sDMvel$ on its own only weakly reduces the scatter at the high-mass end, using it alongside $\ctwohc$ provides a significantly stronger reduction on the scatter than using $\ctwohc$ alone, and this is expected as $\sDMvel$ and $\ctwohc$ are nearly uncorrelated (Figure~\ref{fig:Correlation_Matrix}) and are thus essentially independent properties.

Note that the \textsc{Kllr} scatter for the fiducial result shown in Figure~\ref{fig:SHAP_analysis} will not match that shown in Figure~\ref{fig:sigma_DM_3D_summary} because the halo samples in each are slightly different. In Figure~\ref{fig:SHAP_analysis} we use the concentration in our analysis, and so our halo sample size is reduced by $3 - 4\%$ via the $3\sigma$ outlier rejection performed on this quantity.

The \textsc{Kllr} method and the interpretable ML pipeline have great agreement on the mass-dependence of the property importance --- both predict $\sDMvel$ increasing in importance for larger haloes starting around $10^{12} \msol$, and both predict that $\ctwohc$ is most important at dwarf galaxy-scales and becomes slightly weaker towards cluster-scales. 
An interesting subtlety worth noting is that the \textsc{Kllr} correlations show $\aform$ is strongly correlated with $\sigmaDM$ ($|r| > 0.5$), whereas the ML pipeline says it is uninformative ($\pSHAP < 0.2$) in explaining the scatter. This difference arises because \textsc{Shap} is sensitive to all the different conditional distributions in the problem --- formed by including all possible combinations of the input properties --- whereas our \textsc{Kllr} result is from a distribution that is conditioned on halo mass alone (and is thus marginalized over all the other property dependencies). In the bottom panel of Figure \ref{fig:SHAP_analysis}, we see that conditioning on $\ctwohc$ and $\sDMvel$ (dashed-dotted line) explains $60 - 70\%$ of the total variance, whereas additionally conditioning on all other properties (loosely dotted line) explains only an extra $5-10\%$ of the variance. This is consistent with the picture where $\aform$ correlates strongly with $\sigmaDM$ due to its strong correlations with $\ctwohc$. Thus, once we have conditioned on $\ctwohc$, we find $\aform$ has little importance in explaining the remaining scatter in $\sigmaDM$, as is indicated by \textsc{Shap}.

\section{Discussion} \label{sec:Discussion}

We first summarize results for DMO property scalings, including features common to \textit{both} DMO and FP runs, in section \S \ref{sec:Summary_DMO_Results}, and results for the effects of baryons in \S \ref{sec:Summary_Baryon_Effects}. For both, we also highlight redshift trends detailed in Appendix \ref{appx:Redshift_evol}. In \S \ref{sec:ImpactOfResolution} we discuss the impact of resolution on our results and in \S \ref{sec:UtilityOfKLLR}, we highlight the relevance of \textsc{Kllr} in providing flexible modelling of astrophysical/cosmological observables. Finally, in \S \ref{sec:Sim_Extension}, we discuss possible extensions of our work to other simulation suites.

\subsection{Summary: DMO Property Scaling} \label{sec:Summary_DMO_Results}

The \textsc{Kllr} method offers insight into mass-dependent features in the scaling relations and provides population-level tests of numerical convergence in cosmological simulations. We find that properties of the halo population in a DMO treatment obey simple scaling behaviors with some subtle features that include the following:

\begin{itemize}
    \item The slopes of scaling relations at $z=0$ generally show weak dependence on mass.   The $\ctwohc - \Mtwohc$ DMO relations have a nearly constant slope, $\alpha = -0.10 \pm 0.01$, while the slope of the $\sigmaDM - \Mtwohc$ DMO relations is close to the virial expectation of $\alpha = 1/3$ at high masses but slightly declines (to $\! \sim 0.325$) for haloes less massive than $10^{13} \msol$ (Figures \ref{fig:sigma_DM_3D_summary} and \ref{fig:c200c_summary}).
    \vspace{4pt}
    
    \item The fractional scatters in velocity dispersion and concentration are nearly constant above $10^{11} \msol$, reaching 4\% and 25\%, respectively, and both increase towards lower halo masses. The scatter in halo formation time declines uniformly with halo mass while that of halo shapes increases uniformly.
    \vspace{4pt}
    
    \item Elements of the property correlation matrix tend to be weakly mass-dependent, with the exception of those involving the isotropic velocity dispersion (Figure \ref{fig:Correlation_Matrix}).  The highest mass-dependent correlation values, $r > 0.5$, are seen in the $\sigmaDM - \ctwohc$ and $\sDM - \sDMvel$ pairings. The correlations are generally weakly dependent on redshift (Figure \ref{fig:Redshift_Correlations}).\vspace{4pt}
    
     \item Both \textsc{Kllr} and the interpretable ML method --- which is based on \textsc{XGBoost} and \textsc{Shap} --- show that $\ctwohc$ and $\sDMvel$ are strongly correlated with $\sigmaDM$ and, more importantly, also agree on the specific mass-dependence of this correlation (Figure~\ref{fig:SHAP_analysis}). However, while \textsc{Kllr} also finds a high correlation for $\aform$, the ML method finds none, and this discrepancy arises from differences in the conditional distributions probed by each approach.
     \vspace{4pt}
      
     \item There is generally good numerical consistency in scaling parameters from different resolutions, with most properties showing only mild resolution dependence at the low mass end, where haloes are resolved by $\approx 1500$ particles.
     \vspace{4pt}
     
     \item The redshift evolution of the mean $\sigmaDM - \Mtwohc$ relation is well captured at high masses by the $E(z) = H(z)/H(z = 0)$ factor, where $H(z)$ is the Hubble parameter, but this scaling becomes inaccurate below MW-scales (Figure \ref{fig:Ez_comparison}).
     
\end{itemize}

\subsection{Summary: Baryon Effects of DM Property Scaling} \label{sec:Summary_Baryon_Effects}

The complex hydrodynamic and thermodynamic processes that drive galaxy formation impose mass- and redshift-dependent gravitational forcing on the DM that is not captured in the DMO treatment. By comparing DM halo property relations in the FP and DMO treatments, we find that persistent ``wiggle''-like deviations are introduced into the statistics of the FP halo population, with deviations maximized at the MW mass-scale of $10^{12} \msol$. The specific impact of the FP treatement on DM property scalings include the following:

\begin{itemize}

    \item Haloes near the MW mass scale tend to have higher concentrations and velocity dispersions and are also rounder than their DMO counterparts --- all signals consistent with adiabatic contraction. The amplitudes of these differences can be up to 25\% and also depend on numerical resolution.
    \vspace{4pt}
    
    \item The $\ctwohc$ relation is inverted in the sub-MW mass range of $10^{10.5} \msol < \Mtwohc < 10^{11.5} \msol$; mean concentration \textit{increases} with halo mass, and the trend becomes steeper at higher redshifts (Figure~\ref{fig:c200c_RedEvol}). This inversion is also apparent when conditioning on central galaxy stellar mass, $\Mstarcen$, rather than halo mass, opening up opportunities to observationally search for these features (Section \ref{sec:Observations}). \vspace{4pt}
    
    \item MW-scale haloes form $\approx 0.4 \gyr$ earlier in the FP runs compared to the DMO runs, while haloes of mass $1\,\rm dex$ below this scale tend to form $\approx 0.2 \gyr$ later (Figure~\ref{fig:a_form_summary}). These time delays are consistent with features found in the mass accretion rate, where MW-scale haloes in the FP runs accrete more slowly than their DMO counterparts (Figure~\ref{fig:M_acc_dyn_summary})\vspace{4pt}
    
    \item The FP treatment makes haloes rounder in density-space and velocity-space. This difference is most prominent around $10^{12} \msol$ and becomes increasingly subdued towards dwarf galaxy- and cluster-scales (Figure \ref{fig:shapes_summary}). The mean velocity-space shape also asymptotes to $\sDMvel \approx 0.9$ for $\Mtwohc < 10^{12} \msol$.\vspace{4pt}
    
    \item The deviations in the scaling relations due to the FP treatments arise from the interplay between different galaxy formation processes (Figure~\ref{fig:Feedback_Variation}).  At mass-scales where feedback wins, the DM is displaced and so both concentration and velocity dispersion --- which are sensitive to the halo potential --- are lowered. When cooling and star formation win, the properties both increase.
    \vspace{4pt}
    
    \item The MW-scale feature found in all the relations is associated with the SMBH feedback model. Changing the pivot mass parameter in this model, equation \eqref{eqn:BH_Accretion_Mode}, increases/decreases the average halo mass $\Mtwohc$ at which the SMBH switches from quasar to radio mode feedback, and the MW-scale feature is seen to correspondingly shift its location in mass as well (Figure \ref{fig:Feedback_Variation}).\vspace{4pt}
    
    \item The deviations introduced by the FP treatment have non-trivial redshift dependence, with the amplitude of the MW-scale feature generally decreasing for $\sigmaDM$ and increasing for $\ctwohc$ towards higher redshifts (Figure~\ref{fig:Wiggle_Redshift_evol}).\vspace{4pt}
        
    \item The structure of the property correlation matrix is mostly robust against the introduction of baryonic physics (Figure~\ref{fig:Correlation_Matrix}). The covariance matrix is less robust because the scatter in each property is often sensitive to physics treatment.  

\end{itemize}

\subsection{Impact of Resolution Effects} \label{sec:ImpactOfResolution}

We find clear resolution dependencies in all the FP relations and a few of the DMO ones as well. Here we discuss the interpretation of the baryonic imprints we quantify in this work in light of these resolution-dependent differences.

First, on the galaxy formation side, the \textsc{IllustrisTNG} realizations all share the same astrophysical model parameters, i.e. there is no re-tuning for each resolution level. As was discussed before, this leads to an increase in SFE (and stellar mass) with increasing resolution. The numerical predictions from TNG50 FP --- the highest resolution run in the TNG suite --- are expected to be closer to the ``truth'' compared to those from the TNG100/300 FP runs. However, we stress that for the purpose of qualitatively studying the mass-dependent features in the FP-derived relations, the results from all three runs are equally valuable and highly complementary to one another. Note that one could always consider re-scaling the TNG100/300 results to remove the resolution difference, and such procedures have been employed in many studies of galaxies from the TNG simulations \citep{Pillepich2018FirstGalaxies, Vogelsberger2018MetalsICM, Vogelsberger2020LFfromTNG, Engler2021SMHMrelation}.

Secondly, on the DMO side, the scaling relations are normally well-converged across all three runs for most of the halo mass range. One exception is $\sigmaDM$, whose slopes at all halo masses decrease very slightly with increases in resolution. Other than that, all three of $\ctwohc$, $\sDM$, and $\sDMvel$ show resolution impacts at only the lowest halo masses of each run. There is also the issue of whether the DMO solutions themselves represent the ``true'' propery relations. \citet{Mansfield2021SimBias} recently showed that an ensemble of state-of-the-art DMO simulation suites --- each one run separately by a different group --- still do not converge to a single mean relation for many halo properties, with both the significance of non-convergence and the relevant mass range of this effect varying according to property. One can interpret this non-convergence as a bias in the property relations that arises from a particular choice of simulation hyper-parameters (\eg force softening scale, particle mass/counts). Given the DMO and FP runs used in this work share the same hyper-parameter configurations, their scaling relations should also share similar non-convergence biases, and thus the \textit{relative} differences between the two would be insensitive to said biases. While we do not explicitly validate our claim, we have noted in our prior discussions that the resolution-dependence in the wiggle amplitudes comes almost entirely from the FP-derived relations, not the DMO-derived ones.

\subsection{Beyond power-law modelling with \textsc{Kllr}} \label{sec:UtilityOfKLLR}

A common finding in astrophysical/cosmological studies is that scaling relations are well captured by power-law behavior. For certain analyses, the current accuracy requirements may be such that the power-law continues to be an adequate model, whereas in others there may be a need to improve on this. In general, the field of astronomy as a whole is experiencing an exponential increase in data, and this is true for a vast variety of objects --- stars, exoplanets, galaxies etc. The statistical power provided by this increase will demand that future analyses use more precise models to describe the data.

\textsc{Kllr}, and other similar local linear regression models, provides a powerful yet simple way of moving beyond power-law forms as it models the scaling relation as being \textit{locally} log-linear and with familiar parameters --- normalizations and slopes --- that are now just functions of some scale variable, such as mass, size, luminosity etc. Thus, \textsc{Kllr} is a data-driven method that is completely agnostic to the specific functional form of the scaling relations, and this is particularly powerful as one need not know apriori what analytic expression best fits the data. 

The \textsc{Kllr} methods are publicly available, as was noted before, and both the formalism and full functionality will be detailed further in a future publication (Farahi, Anbajagane \& Evrard, in prep.). 

\subsection{Extension to other simulation suites} \label{sec:Sim_Extension}

Finally, we address the fact that our results have focused on a single simulation suite. As was shown in Figure~\ref{fig:Feedback_Variation} and discussed in Section \ref{sec:Impact_Feedback}, the locations of the deviations in the FP treatment arise from galaxy formation physics and are sensitive to the specific astrophysical model configurations. Thus, an additional verification step is required, where results from \textit{multiple} simulation suites are compared to one another to test the robustness of our predictions. For example, the \textsc{Magneticum Pathfinder} suite\footnote{\url{https://www.magneticum.org/}} and the \textsc{Bahamas} suite \citep{McCarthy2017BAHAMAS} contain the requisite realizations --- albeit at a lower resolution than \textsc{IllustrisTNG}, which then limits the mass range that can be probed --- to perform similar analyses. In fact, \citet{Henson2017BaryonEffectsBM} have already done comparisons between DMO and FP runs for cluster-scale haloes in the largest box of the \textsc{Bahamas} suite.

The baryonic processes we discuss in this work --- gas cooling and stellar/SMBH feedback --- are standard components of all galaxy formation models. Thus, the features we discuss will generalize to all hydrodynamics simulations, but the precise locations and amplitudes could vary. We note that the study of \citet{Anbajagane2021VelocityBias} compares the $\sigmaDM - \Mtwohc$ relation across an ensemble of hydrodynamics simulations --- which includes \textsc{IllustrisTNG}, \textsc{Magneticum Pathfinder}, and \textsc{Bahamas} --- and consistently finds that the slope decreases to $\alpha < 1/3$ towards galaxy group-scale haloes. This is a promising indication that the qualitative (and possibly quantitative) features we detail here are indeed generalizable to other hydrodynamics simulations. We hope the methodologies and tools presented in this work enable the community to perform similar analyses on other simulations suites and thus enable convergence tests of these predictions.

\section{Conclusions} \label{sec:conclusions}

Halo property scaling relations from simple gravitational evolution of DM have no special scales, but the inclusion of galaxy formation physics introduces many scale-dependent features. We quantify the location and amplitude of these features by extracting the mass-dependent scaling relation parameters --- normalizations, slopes, and covariances --- for a set of five DM halo properties from simulated haloes in both dark matter only (DMO) and full physics (FP) runs of the IllustrisTNG suite. The joint halo sample contains $1.5$ million haloes and covers nearly six decades in mass, $10^{9} \msol < \Mtwohc < 10^{14.5} \msol$, from dwarf galaxies to galaxy clusters.

The scaling relations in the DMO runs are simple as expected, but have some features of interest discussed in \S \ref{sec:Summary_DMO_Results}. The inclusion of galaxy formation physics induces deviations in every scaling relation which persist across all redshifts and can be up to 25\% in amplitude. The specific features are summarized in \S \ref{sec:Summary_Baryon_Effects}. The location and nature of these deviations are connected to the baryonic processes --- gas cooling and stellar/SMBH feedback --- used in the TNG FP model.

The DMO-derived scaling relations for many of the properties studied here are frequently used in the modelling of astrophysical/cosmological observables, and we provide a means to quantify whether the impact of baryonic physics can be ignored given the accuracy requirements of an analysis. Studies of extensions to the $\Lambda \rm CDM$ model, such as self-interacting dark matter (SIDM) and modified gravity (MG), normally use a CDM N-body simulation as the reference model and explore deviations from this reference caused by physics of the model extensions. However, as we show here, the deviations from just baryonic effects can be significant and add additional uncertainties in the reference model as well. These must be taken into account when performing any inference on observational data.

We can also invert the exercise by observationally measuring these scaling relations for MW-scale haloes and above, and then use any observed ``wiggle'' features (or lack thereof) to set constraints on the physics of SMBH feedback mechanisms. This is a particularly exciting prospect as the SMBH feedback is a leading systematic in cosmological studies of DM clustering on small-scales and so any information on the scale of these properties would be useful.

Precise calibration of galaxy formation features is required to debias cosmological studies and empower new physics searches based on DM halo structure. Thus, studies such as the one presented here impact a wide variety of science cases, and so we encourage \textsc{Kllr}-like analyses of other current and future simulation suites and of other additional halo properties of interest to the community.

\section*{Acknowledgements}

We thank Han Aung, Daisuke Nagai, Dylan Nelson, and Annalisa Pillepich for helpful discussions, and Phil Mansfield for providing extensive feedback on an earlier version of this draft. We additionally thank Dylan Nelson and Annalisa Pillepich for providing us with access to the TNG model variant runs which are currently proprietary datasets, and also thank the whole TNG team for publicly releasing all data from their suite of simulations. Finally, we thank Dylan Nelson once more for his help with making our halo catalogs public on the TNG database.

DA is supported by the National Science Foundation Graduate Research Fellowship under Grant No. DGE 1746045. AF is partially supported by a Michigan Institute for Data Science (MIDAS) Fellowship.

All analysis in this work was enabled greatly by the following software: \textsc{Pandas} \citep{Mckinney2011pandas}, \textsc{NumPy} \citep{vanderWalt2011Numpy}, \textsc{SciPy} \citep{Virtanen2020Scipy}, \textsc{Sci-kit Learn} \citep{Pedregosa2012Sklearn}, and \textsc{Matplotlib} \citep{Hunter2007Matplotlib}. We have also used
the Astrophysics Data Service (\href{https://ui.adsabs.harvard.edu/}{ADS}) and \href{https://arxiv.org/}{\texttt{arXiv}} preprint repository extensively during this project and the writing of the paper.

\section*{Data Availability}

The \textsc{Kllr} parameters of all property scaling relations presented in this work are available on GitHub at \url{https://github.com/DhayaaAnbajagane/Baryon_Imprints_TNG}. This includes the halo mass-dependent means, slopes, scatter, skewness, and correlations. These data products are available for all six TNG runs at twenty redshifts between $0 < z < 12$. We also provide a convenience script that parses the scaling parameter data files while also being able to interpolate between halo mass and redshift as needed.

The raw halo property catalogs used in deriving the scaling relations are available on the TNG website\footnote{\url{http://www.tng-project.org/anbajagane21}} as supplementary catalogs which complement the existing, extensive public data releases described in \citet{Nelson2019TNGPublicData}. Our supplementary halo catalogs also include additional properties not discussed in this work.




\bibliographystyle{mnras}
\bibliography{References} 




\appendix

\section{Additional Properties} \label{appx:Additional_Properties}

In addition to the properties explored in the main text, we also explore two more properties here: (i) the mass accretion rate of the halo, $\Gammadyn$, which captures the recent growth history, and (ii) the energy content of the DM surface pressure $\EsDM$, which captures the current dynamical state. For each property, we present their definition first, and then their \textsc{Kllr} parameters for $z = 0$. Their correlations with all other parameters are also shown alongside the redshift evolution analysis (Figure~\ref{fig:Redshift_Correlations}).

\subsection{Mass accretion rate, \texorpdfstring{$\Gammadyn$}{M\_acc\_dyn}} 
\label{appx:M_acc_dyn}

\subsubsection{Definition}

The mass accretion rate (MAR) is the logarithmic derivative of the halo mass, $\Mtwohc$ with respect to the cosmic scale factor, $a$, averaged over a specified time scale. Here, the time scale is the dynamical time \citep[see eqn 5]{Diemer2017SPARTA}. The MAR is then formally defined as,
\begin{equation} \label{eqn:MAR}
    \Gammadyn = \frac{\ln M(a_{\rm now}) - \ln M(a_{\rm dyn})}{\ln a_{\rm now} - \ln a_{\rm dyn}},
\end{equation}
where $a_{\rm dyn}$ is the cosmic scale factor exactly one dynamical time, $\tdyn$, ago from the present. This dynamical time, which is also the halo crossing time, can be computed for a density contrast $\Delta = 200c$ as
\begin{equation} \label{eqn:tdyn}
    \tdyn = \frac{1}{5H(a_{\rm now})},
\end{equation}
where $H(a)$ is the Hubble parameter. At the present epoch ($a = 1$), $\tdyn \approx 2.9 \,\gyr $, and so $a_{\rm dyn} \approx 0.81$.

In practice, we compute $\Gammadyn$ by first constructing a data vector of the halo mass $\Mtwohc(\ln a)$ at various snapshots (scale factors) by following the subhalo merger tree of the primary, or ``central'' subhalo of the given halo. We then interpolate linearly over the function $\ln \Mtwohc (\ln a)$ to get $\Mtwohc(a_{\rm dyn})$, and use equation \eqref{eqn:MAR} to obtain $\Gammadyn$. The merger trees were computed using \textsc{SubLink} \citep{Rodriguez-Gomez2015SubLinkTrees}. Once again, the merger tree follows the central subhalo evolution, but $\Gammadyn$ is defined by $\Mtwohc$, which is computed using all available particles (section \ref{sec:mass_def}).

Alternative approaches compute a parametric mass accretion \textit{history}, $dM/dt$, which models the \textit{smooth} accretion of a halo over its cosmic history \citep[\eg][]{Wechsler2002Concentrations, Zhao2003MahAndConcentration, Tasitsiomi2004ModelMAH, McBride2009ModelMAH, Fakhouri2010ModelMAH, Hearin2021DiffMAH}. However, such models by construction do not capture time-transient accretion features --- primarily arising from merger events but also from baryonic processes, like outflows, which are the focus of this work --- and so we do not employ this framework here.

\subsubsection{LLR Fit and Parameters}

\begin{figure}
    \centering
    \includegraphics[width = \columnwidth]{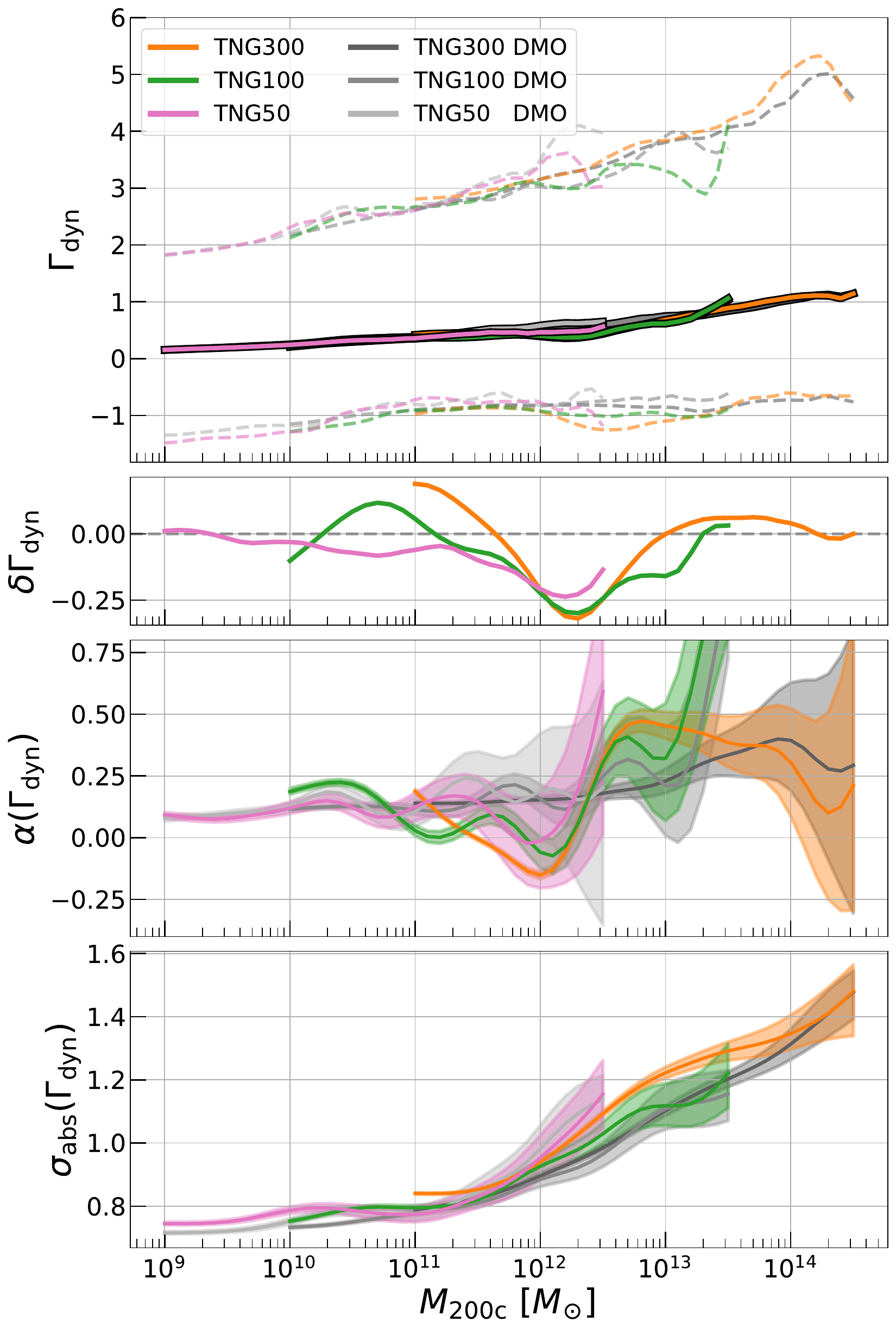}
    \caption{The $z = 0$ \textsc{Kllr} scaling parameters of the mass ccretion rate $\Gammadyn$ relation. We use the same style conventions as Figure~\ref{fig:sigma_DM_3D_summary}, with two expections: the upper middle panel shows $\delta \Gammadyn = \Gammadyn^{\rm FP}/\Gammadyn^{\rm DMO} - 1$, and the scatter is an absolute scatter, \textit{not} a fractional one. The mean $\Gammadyn$ is positive at all mass-scales, and increases with mass. In the FP runs, $\Gammadyn$ is $25\%$ lower at MW-mass scales alone. The spread of $\Gammadyn$ is strongly skewed at high halo mass.}
    \label{fig:M_acc_dyn_summary}
\end{figure}

Since $\Gammadyn$ is a \textit{signed} variable and can have negative values, we do not take the logarithm of this quantity. The extracted scatter is then an absolute scatter and not a fractional one.

The $\Gammadyn - \Mtwohc$ mean relation (top panel, figure \ref{fig:M_acc_dyn_summary}) increases with halo mass for all runs. The DMO run mean relations are within $10 - 20\%$ of the fitting function from \citet{Diemer2017SplashbackIIonMARs} for all halo masses (figure not shown), though a precise comparison is not possible given differences in halo mass and accretion rate definitions between the two works. 

In the FP runs, $\Gammadyn$ is suppressed by $25\%$ at MW-scales, and this could be due to feedback overpowering the shallower halo potential and generating significant outflows which lead to a lower accretion rate \citep[\eg][]{Croton2006AGNFeedback, Schaye2015EAGLE, Muratov2015GasOutflowsFIRE}. The slopes and scatter are insensitive to resolution and baryonic physics, with the exception of the MW-scale feature in the slopes. Note that the lower $\Gammadyn$ at MW-scales in the FP runs is connected to results in Figure~\ref{fig:a_form_summary}, where we show the formation time of the MW-scale haloes in the FP runs is also earlier than their DMO counterparts. Since the FP haloes have a lower accretion rate, they grow more slowly, and thus need to form at an earlier epoch in order to have enough time to accumulate matter and become a MW-scale halo at the present epoch.

\subsection{DM surface pressure energy, \texorpdfstring{$\EsDM$}{EsDM}}
\label{appx:Es_DM}

\subsubsection{Definition}

The DM surface pressure energy, $\EsDM$, captures the relaxation state of the halo and is a required additional term in the virial relation when describing haloes as the halo mass profiles do not have a sharp, finite boundary and instead gradual transition into the mean, background field. We follow the same prescription from \citet{Shaw2006StatisticsClusters} for computing this property,

\begin{align} \label{eqn:Shaw_EsDMO_definitions}
    P_s & = (1/3)\frac{\sum_i m_i v_i^2}{V}\\
    V & = \frac{4}{3}\pi\Rtwohc^3 - \frac{4}{3}\pi R_{0.8}^3\\
    E_s &\approx 4\pi R_{0.9}^3 P_s,
\end{align}

where $m_i$ is the mass of the DM particle, $v_i$ is the magnitude of the relative velocity with respect to the mean velocity of all DM particles within $\Rtwohc$. Finally, $R_N$ with $N \in [0, 1]$ is the radii of a spherical shell that contains a fraction $N$ of our DM particle set, which is defined as all particles bound to the central subhalo while within a distance $\Rtwohc$ from it. We estimate $R_N$ by rank-ordering the particles by their distance from the halo center, and then picking the $N^{\rm th}$ quantile. The sum in equation \eqref{eqn:Shaw_EsDMO_definitions} is performed using all DM particles within $R_{0.8} < r < \Rtwohc$.

$\EsDM$ has units of energy and we choose to normalize it to remove this dimensional dependence. For the rest of this section, we refer to the quantity $\EsDM/(V_{\rm 200c}^2\Mtwohc)$ instead, where
\begin{equation}\label{eqn:V200c}
    V_{\rm 200c} = \sqrt{\frac{G\Mtwohc}{\Rtwohc}} \propto (\Mtwohc)^{1/3},
\end{equation}
is the velocity scale of the halo and depends on halo mass.

\subsubsection{LLR Fit and Parameters}

\begin{figure}
    \centering
    \includegraphics[width = \columnwidth]{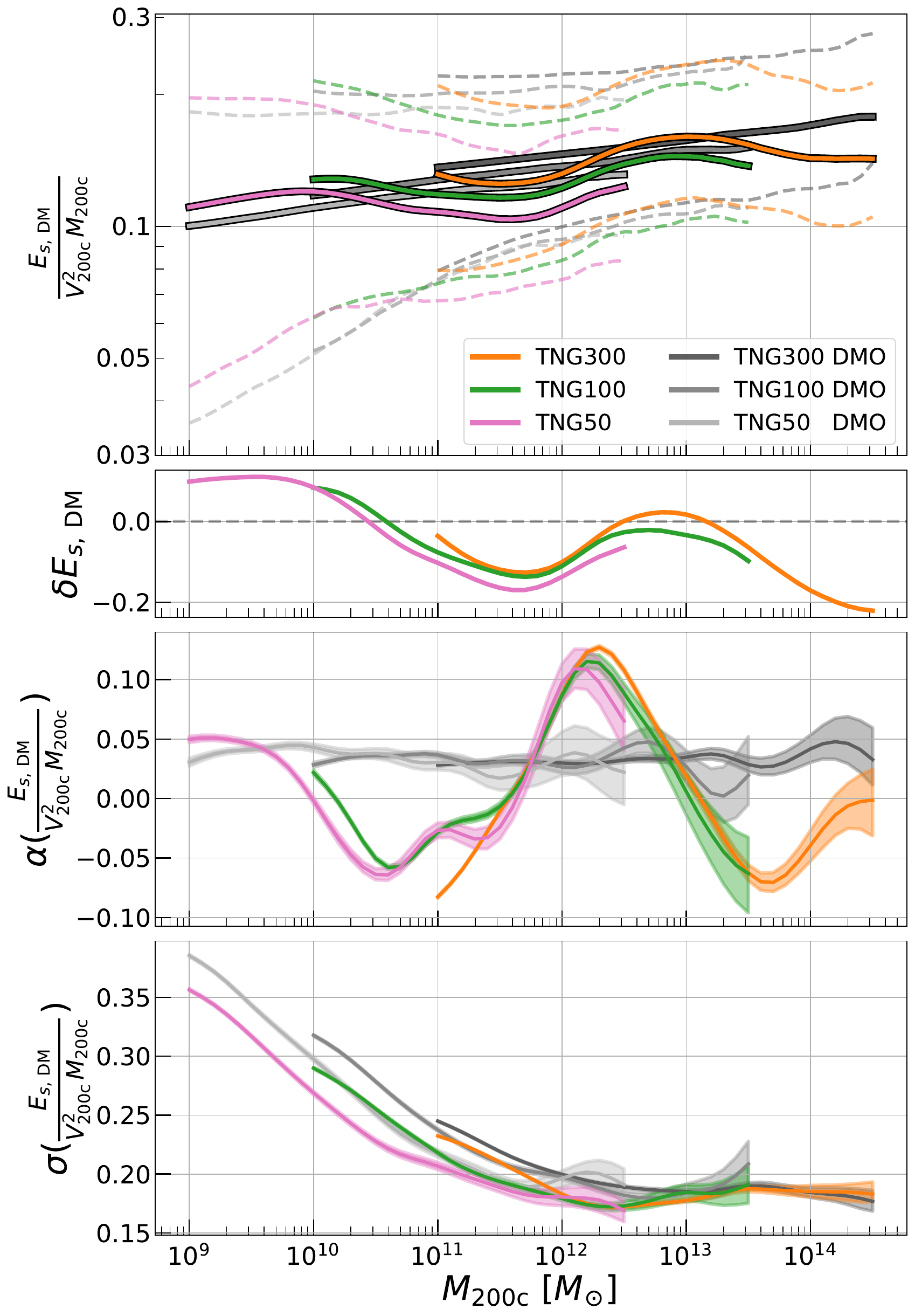}
    \caption{The scaling relation (top), slope (middle), and scatter (bottom) of the $\EsDM - \Mtwohc$ relation, but with $\EsDM$ normalized to remove dimensional dependence. We find that the mean $\EsDM$ is consistently lower than $20\%$ of the characteristic kinetic energy scale of the halo, $V_{\rm 200c}^2 \Mtwohc$. The slope of the FP runs is non-monotic but also changes sign \textit{thrice} over this mass range, whereas the DMO runs maintain a constant slope of $\alpha \approx 0.04$.}
    \label{fig:EsDMO_Summary}
\end{figure}

The $\EsDM - \Mtwohc$ mean relations show that halo surface pressure energy tends to be $10 - 15\%$ the characteristic kinetic energy scale (Figure~\ref{fig:EsDMO_Summary}). The relations also have clear dependencies on resolution and baryonic physics. The latter is expected as $\EsDM$ depends on both the DM velocity dispersion and halo concentration, both of which show strong features in the FP runs. The resolution impact in \textit{both} DMO and FP runs is less expected as in the DMO runs $\sigmaDM$ has only a subtle dependence on resolution. However, the surface pressure probes the velocity dispersion \textit{profile} rather than the integrated quantity given it is only sensitive to the outer shell of the halo, and we do find a resolution dependence in the velocity dispersion computed in such a shell (figure not shown). Given $\EsDM$ depends on $\ctwohc$ and $\sigmaDM$, both of which have nearly constant slopes in the DMO runs, it is no surprise that the DMO runs here also have a constant slope ($\alpha \approx 0.04$).

The mass-dependent features for the FP runs (in comparison to DMO relations) go as follows: the surface pressure is higher at dwarf-galaxy scales, lower at MW-scales, approximately the same at galaxy group-scales, and lower again at cluster-scales. The slopes reflect this complicated behavior, as they show nearly four different zero-crossings corresponding to minima/maxima in the mean relation. \citet{Cui2017DynamicalState} studied the same quantity for cluster-scale haloes ($\Mtwohc > 10^{14.5} \msol$) and found evidence that $\EsDM$ is lowered by the inclusion of baryons. They also see mass-dependent deviations between their FP and DMO runs.

To our knowledge, this is the first time this property scaling relation has been extracted for the six decades in mass probed here.

\section{Redshift Evolution} \label{appx:Redshift_evol}

In section \ref{sec:Impact_Feedback}, we showed that the differences between the FP and DMO runs exist, and are significant, at higher redshifts as well. Here, we explore the redshift evolution of the $\sigmaDM - \Mtwohc$ relation, the $\ctwohc - \Mtwohc$ relation, and the correlation matrix. To prevent overcrowding in our plots, we show only TNG300 and TNG50, a combination that still probes the entire halo mass range of the main analysis. We also stress again that results for $0 < z < 12$ are provided in our public release of the scaling relation parameters.

\subsection{DM velocity dispersion, \texorpdfstring{$\sigmaDM$}{sigmaDM}} \label{appx:RedEvo_sigmaDM}

\begin{figure}
    \centering
    \includegraphics[width = \columnwidth]{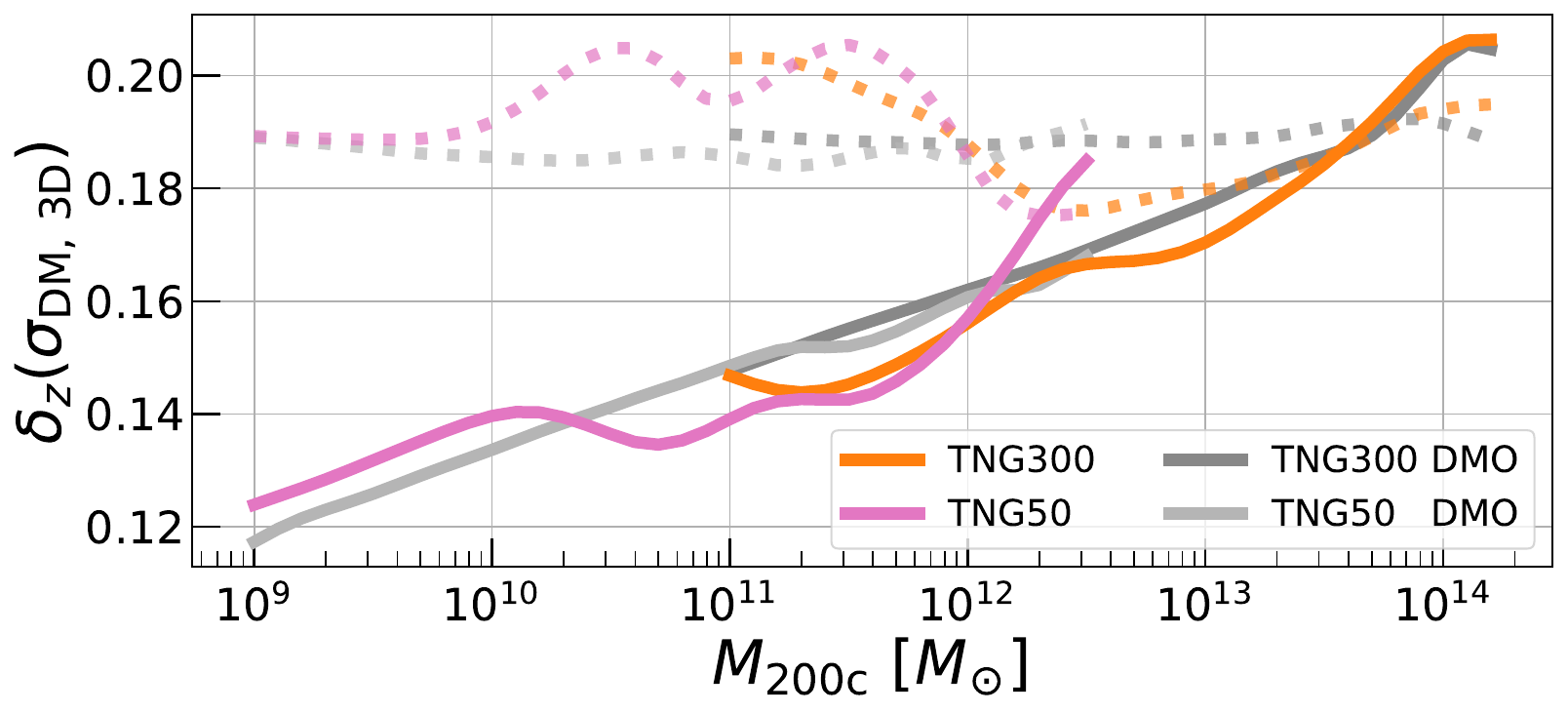}
    \caption{the fractional deviations between the mean $\sigmaDM - \Mtwohc$ relations of the different redshift runs, $\delta_z = \ln\sigmaDM(z = 1) - \ln\sigmaDM(z = 0)$. The dotted lines show $\ln [E(z)^\alpha]$, which is the expected redshift scaling from \citet{Evrard2008VirialScaling}. This scaling works for high mass haloes but is a poorer description for the low halo-mass end.}
    \label{fig:Ez_comparison}
\end{figure}

Figure~\ref{fig:Ez_comparison} studies the redshift evolution of the $\sigmaDM - \Mtwohc$ mean relation. Previous work has shown that for haloes of $\Mtwohc > 10^{14} \msol$, the redshift evolution of the $\sigmaDM - \Mtwohc$ mean relation can be captured by scaling the halo mass as $\Mtwohc \rightarrow E(z)\Mtwohc$, where $E(z) = H(z)/H_0$ \citep{Evrard2008VirialScaling, Singh2020CosmologyWithMORs}. This implies that the higher redshift scaling relations can be predicted as
\begin{equation}\label{eqn:sigmaDM_E08_scaling}
    \ln \sigmaDM(z, M) \approx \ln \sigmaDM(z = 0, M) + \alpha(z = 0, M) \ln E(z),
\end{equation}
where both $\ln \sigmaDM$ and $\alpha$ are evaluated locally at mass scale, $M$.

However, Figure~\ref{fig:Ez_comparison} shows that while this $E(z)$ scaling is still appropriate for the high-mass haloes, it consistently predicts a larger $\sigmaDM$ --- by up to $7\%$ --- for the rest of the halo mass range. This mass-dependence arises because the slopes have a redshift dependence that is not incorporated in equation \eqref{eqn:sigmaDM_E08_scaling}.

\subsection{NFW Concentration, \texorpdfstring{$\ctwohc$}{c200c}} \label{appx:RedEvo_c200c}

The redshift evolution of the $\ctwohc - \Mtwohc$ relation has been studied extensively in DMO simulations \citep[\eg][]{Bullock2001Concentrations, Wechsler2002Concentrations, Duffy2008Concentration, Bhattacharya2013Concentrations, Ludlow2014MilleniumConcentration, Klypin2016Concentrations, Diemer2015Concentration, Correa2015Concentration, Ludlow2016Concentration, Child2018ConcentratioMassRelation, Diemer2019concentrations, Ishiyama2020UchuuConcentration}, and it is known that the slope of the relation changes significantly over redshift.

\begin{figure}
    \centering
    \includegraphics[width = \columnwidth]{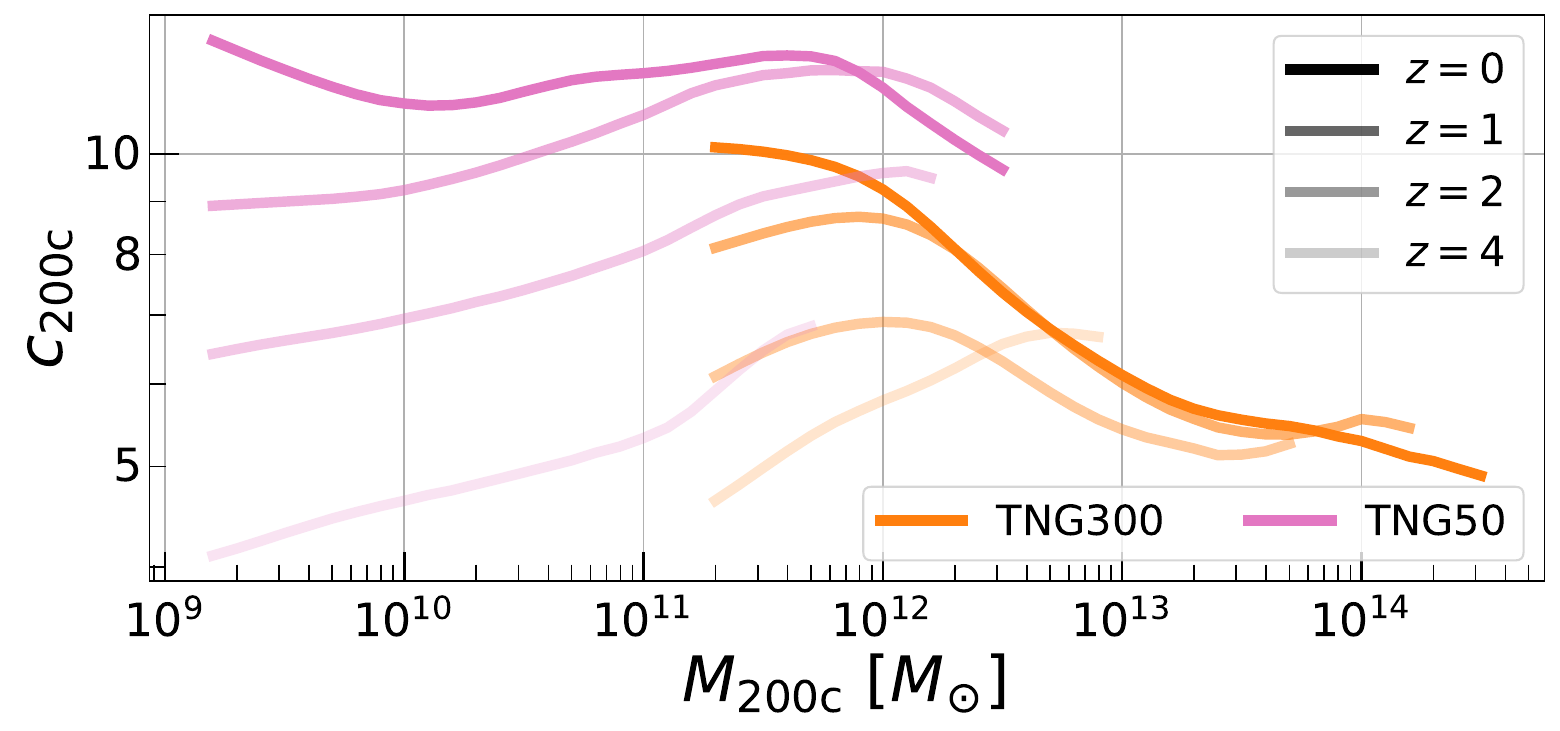}
    \caption{The $\ctwohc - \Mtwohc$ relation at different redshifts for the FP runs. For dwarf-galaxy scale haloes at higher redshifts, $\ctwohc$ \textit{increases} with $\Mtwohc$, and this behavior becomes steeper with increasing redshift. At the same mass- and redshift-scales, the DMO run relations (not shown here) continue exhibiting a negative slope.}
    \label{fig:c200c_RedEvol}
\end{figure}

Interestingly, Figure~\ref{fig:c200c_RedEvol} shows that in the FP runs the slopes become \textit{positive} as early as $z = 1$. This is most prominent for dwarf galaxy-scale haloes, though MW-scale objects also share this behavior at high redshift. Notably, at the mass- and redshift-scales where the FP runs show a positive slope, the DMO runs still exhibit a negative slope. This implies that the presence of baryonic physics has flipped the sign of the slope.

\subsection{Correlation Matrix} \label{appx:Correlation_Matrix}

\begin{figure*}
    \centering
    \includegraphics[width = 2\columnwidth]{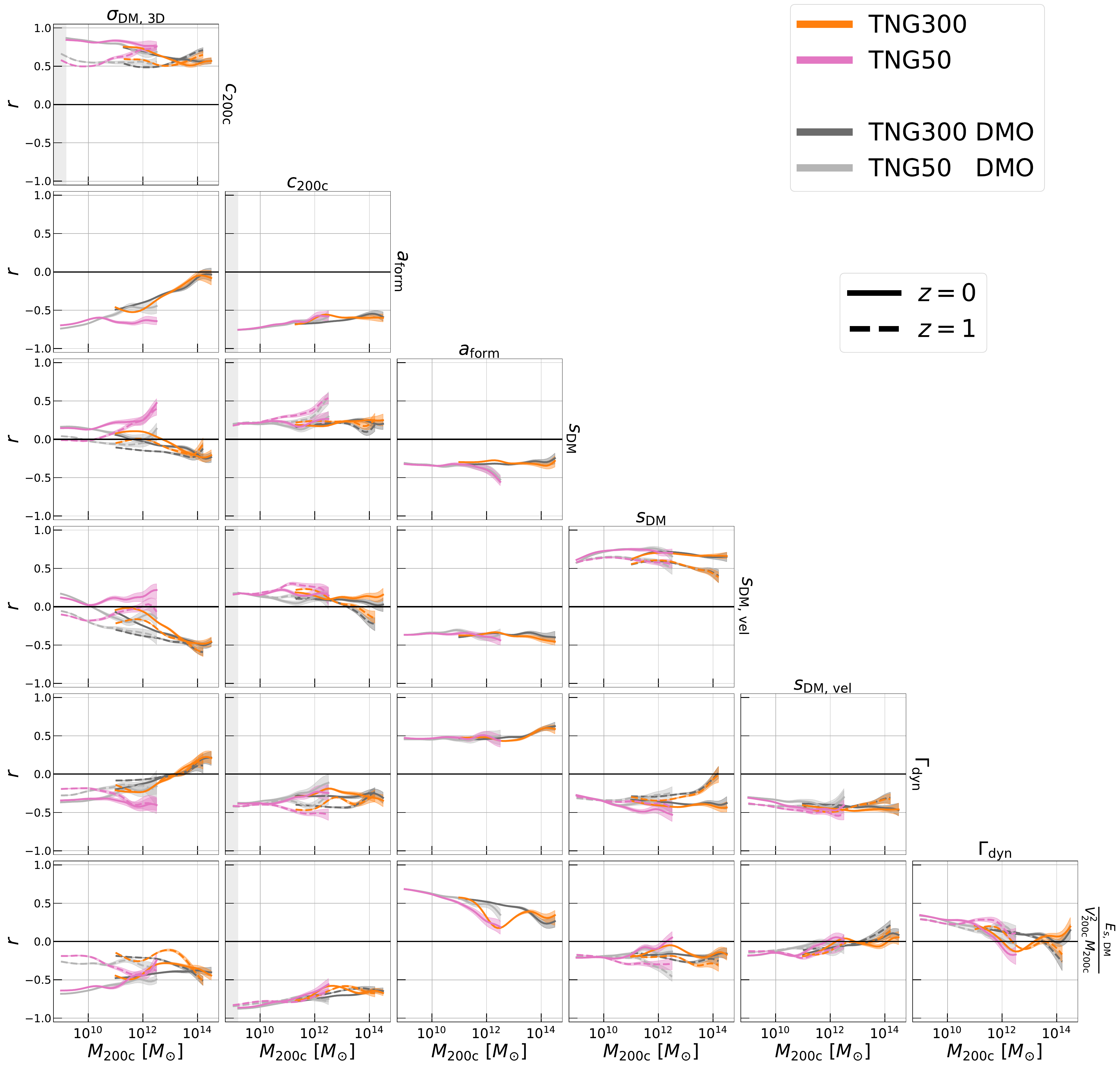}
    \caption{The full \textsc{Kllr} correlation matrix for all seven parameters, including the two introduced in the appendix, for both $z = 0$ (solid line) \textit{and} $z = 1$ (dashed line). We omit $\aform$ from the $z = 1$ analysis due the ambiguity in defining two different formation times for a single halo (see text for details). We also do not show TNG100 to enhance visibility within subplots, and because the TNG100 results are always bounded by the TNG50 and TNG300 results.}
    \label{fig:Redshift_Correlations}
\end{figure*}

The full correlation matrix in Figure~\ref{fig:Redshift_Correlations} shows some relatively small redshift evolution in the correlations. The $\sigmaDM - \ctwohc$ correlation undergoes the largest change, as it is now mass-independent at $r \approx 0.5$. The shape parameters $\sDM - \sDMvel$ are less correlated but still have $r > 0.5$. We do not compute $\aform$ for halo catalogs at $z > 0$ as it is ambiguous to assign a single halo multiple formation times. Thus, any correlations with $\aform$ at $z = 1$ are omitted from Figure~\ref{fig:Redshift_Correlations}.

\bsp	
\label{lastpage}
\end{document}